\documentclass[a4paper]{article}
\usepackage{jheppub}
\usepackage[utf8]{inputenc}
\usepackage{tikz}
\usetikzlibrary{cd}
\usetikzlibrary {matrix}
\usetikzlibrary{shapes,arrows,backgrounds,fit,positioning}

\usepackage{calc}

\newcommand\refr[1]      {ref.\,\cite{#1}}
\newcommand\refrs[1]    {refs.\,\cite{#1}}

\newcommand\eqn[1]     {eq.\,(\ref{#1})}
\newcommand\eqns[2]    {eqs.\,(\ref{#1}) and~(\ref{#2})}
\newcommand\eqnss[2]   {eqs.\,(\ref{#1})--(\ref{#2})}

\newcommand\fig[1]     {figure~{\ref{#1}}}

\newcommand\sect[1]    {section~{\ref{#1}}}

\def\beq{\begin{equation}}
\def\eeq{\end{equation}}
\def\bsp#1\esp{\begin{split}#1\end{split}}
\def\bal#1\eal{\begin{align}#1\end{align}}

\newcommand\tgam[1]    {\Gamma^{\mathrm{#1}}}
\newcommand\btgam[1]   {\overline{\Gamma}^{\mathrm{#1}}}
\newcommand\dgam[1]    {\rd\Gamma^{{\rm #1}}}
\newcommand\dgama[2]   {\rd\Gamma^{{\rm #1,A}_{\scriptscriptstyle #2}}}

\newcommand\la         {\langle}
\newcommand\ra         {\rangle}
\newcommand\bra[3]     {\la {\cal M}_{#1}^{#2}#3|}
\newcommand\ket[3]     {|{\cal M}_{#1}^{#2}#3\ra}
\newcommand\braket[4]  {\la {\cal M}_{#1}^{#2}#4|{\cal M}_{#1}^{#3}#4\ra}
\newcommand\SME[3]     {|{\cal M}_{#1}^{#2}{#3}|^2}

\newcommand{\rd}       {{\mathrm{d}}}
\newcommand{\PS}[1]    {\rd\phi_{#1}}

\newcommand\kT[1]      {k_{\perp#1}}
\newcommand\hz[1]   {\hat{z}_{#1}}
\newcommand\hkT     {\hat{k}_{\perp}}

\newcommand{\cC}[2]    {{\cal C}_{#1}^{#2}}
\newcommand{\cS}[2]    {{\cal S}_{#1}^{#2}}

\newcommand{\cSCS}[2]  {{\cal C}\kern-2pt{\cal S}_{#1}^{#2}}

\newcommand{\rC}       {{\mathrm C}}
\newcommand{\rS}       {{\mathrm S}}
\newcommand{\rSCS}     {{\rC}\kern-2pt{\rS}}
\newcommand{\IcC}[2]   {{\rC}_{#1}^{#2}}
\newcommand{\IcS}[2]   {{\rS}_{#1}^{#2}}

\newcommand{\IcSCS}[2] {\rC\kern-2pt\rS_{#1}^{#2}}
\newcommand{\bI}       {\bom{I}}

\newcommand{\cF}       {{\cal F}}

\newcommand{\cM}       {{\cal M}}
\newcommand{\Li}   {\mathrm{Li}}

\newcommand{\CF}       {C_{\mathrm{F}}}
\newcommand{\CA}       {C_{\mathrm{A}}}
\newcommand{\TR}       {T_{\mathrm{R}}}

\newcommand{\bT}       {\bom{T}}

\newcommand\qb         {{\bar q}}
\newcommand\bb         {{\bar b}}

\newcommand{\gam}	{\ensuremath{\gamma}}

\def\AP{Altarelli--Parisi } 
\newcommand\bom[1]     {{\mbox{\boldmath $#1$}}}

\newcommand{\wha}[1]   {\widehat{#1}}

\newcommand{\ti}[1]   {\widetilde{\,#1\,}}
\newcommand{\ha}[1]   {\widehat{\,#1\,}}

\newcommand\al	{\alpha}
\newcommand\be	{\beta}

\newcommand{\ep}       {\epsilon}

\newcommand\vph {\ensuremath{\varphi}}

\newcommand{\cA}       {{\cal A}}

\newcommand\Oe[1]      {\ensuremath{\mathrm O(\ep^{#1})}}
\newcommand\Oa[1]      {\ensuremath{\mathrm O(\as^{#1})}}

\newcommand\as  	       {\ensuremath{\alpha_{\mathrm{s}}}}

\newcommand\MSbar	{\ensuremath{\overline{\mathrm{MS}}}}

\def\hP			      {\hat{P}}

\def \Li{{\rm Li}}

\def \aih{\alpha_{\ha{i}}}
\def \ajh{\alpha_{\ha{k}}}
\def \dih{d_{\ha{i}}}
\def \djh{d_{\ha{k}}}
\def \dijh{(\dih+\djh)}
\def \vmh{\ha{v}_-}
\def \vph{\ha{v}_+}
\def \lvh{\ln(\ha{v})}
\def \lvph#1{\ln^{#1}(\ha{v}_+)}
\def \lnih#1{\ln^{#1}\Big(\frac{\aih}{\vph}\Big)}
\def \lniph#1{\ln^{#1}\Big(1-\frac{\aih}{\vph}\Big)}
\def \lnjh#1{\ln^{#1}\Big(\frac{\ajh}{\vph}\Big)}
\def \lnjph#1{\ln^{#1}\Big(1-\frac{\ajh}{\vph}\Big)}
\def \Li{{\rm Li}}
\def \liih#1{{\Li_#1\Big(\frac{\aih}{\vph}\Big)}}
\def \lijh#1{{\Li_#1\Big(\frac{\ajh}{\vph}\Big)}}
\def \lixh#1{{\Li_#1(\ha{x})}}
\def \lixph#1{{\Li_#1(1-\ha{x})}}

\title{Fully exclusive heavy quark-antiquark pair production from a colourless initial 
state at NNLO in QCD}

\author[a]{G\'abor Somogyi}
\author[b]{and Francesco Tramontano}

\affiliation[a]{MTA-DE Particle Physics Research Group, 
    University of Debrecen, 4010 Debrecen, PO Box 105, Hungary} 
\affiliation[b]{Universit\`a di Napoli and INFN, Sezione di
  Napoli, Complesso Universitario di Monte Sant'Angelo,
  Via Cintia, 80126 Napoli, Italy} 

\emailAdd{somogyi.gabor@science.unideb.hu}
\emailAdd{francesco.tramontano@unina.it}

\abstract{We present a local subtraction scheme for computing 
next-to-next-to-leading order QCD corrections to the production 
of a massive quark-antiquark pair from a colourless initial state. 
The subtraction terms are built following the CoLoRFulNNLO method 
and refined in such a way that their integration gives rise to compact, 
fully analytic expressions. 
All ingredients necessary for a numerical implementation of our 
subtraction scheme are provided in detail. As an example, 
we calculate the fully differential decay rate of the Standard Model 
Higgs boson to massive bottom quarks at next-to-next-to-leading order 
accuracy in perturbative QCD. 
}
\date{July 2020}

\begin{document}

\maketitle

\section{Introduction}
\label{sec:intro}

Actual and planned CERN LHC operation opens the possibility
to perform a large number of accurate measurements in high 
energy physics. It is clear that for many of them the overall 
experimental uncertainty will be much smaller than the theory
uncertainty estimated based on next-to-leading order
QCD corrections. Then, including higher order corrections 
turns out to be mandatory for a meaningful comparison among theory
predictions and experimental data. 
Thus, next-to-next-to-leading order (NNLO) computation has received 
considerable attention and several approaches for performing these 
calculations have been proposed. These include the $q_T$~\cite{Catani:2007vq}
and $N$-jettiness~\cite{Boughezal:2015dva,Gaunt:2015pea} slicing methods, 
as well as subtraction schemes, like antenna~\cite{GehrmannDeRidder:2005cm,
Daleo:2006xa,GehrmannDeRidder:2005aw,GehrmannDeRidder:2005hi,Daleo:2009yj,
Gehrmann:2011wi,Boughezal:2010mc,GehrmannDeRidder:2012ja,Currie:2013vh},
CoLoRFulNNLO~\cite{Somogyi:2005xz,Somogyi:2006da,Somogyi:2006db,Somogyi:2008fc,
Aglietti:2008fe,Somogyi:2009ri,Bolzoni:2009ye,Bolzoni:2010bt,DelDuca:2013kw,
Somogyi:2013yk}, residue-improved~\cite{Czakon:2010td,Czakon:2011ve,
Czakon:2014oma,Czakon:2019tmo}, nested soft-collinear~\cite{Caola:2017dug,
Caola:2018pxp,Delto:2019asp,Caola:2019nzf,Caola:2019pfz} and 
projection-to-Born~\cite{Cacciari:2015jma} with yet other 
approaches under development~\cite{Magnea:2018hab,Magnea:2018ebr,Herzog:2018ily}. 

From the mathematical point of view, computations at NNLO  
are more elaborate than ones at NLO and for this reason the 
level of automation is still much less advanced.
On the one hand, difficulties lie in the computation of the double 
virtual amplitudes for processes with many particles in the final state
and with masses. On the other hand, there are still no fully 
satisfactory, complete and general algorithms for the regularization 
of the infrared divergences for fully differential NNLO computation 
as there are for the NLO case. Here by general, we refer to a scheme
that applies to any kind of singularity coming from both initial 
and massive or massless final state particles.
By complete, we mean that the set of subtractions defining a scheme 
is given in full detail together with the complete integrated versions. 
This last point is of course mandatory to allow for independent applications, 
validations or just implementations for analysis purposes.

Here we consider the CoLoRFulNNLO method that has been formulated
in \refr{Somogyi:2005xz,Somogyi:2006da,Somogyi:2006db,Somogyi:2008fc,
Aglietti:2008fe,Somogyi:2009ri,Bolzoni:2009ye,Bolzoni:2010bt,DelDuca:2013kw,
Somogyi:2013yk}. The basic ideas of the method are quite general
and apply to final state as well as initial state singularities.
We focus on the final state infrared singularities and in particular, 
on the production of a massive quark-antiquark pair from a colourless 
initial state. NNLO corrections to such processes have been computed 
in the literature previously \cite{Gao:2014eea,Gao:2014nva,Chen:2016zbz,
Bernreuther:2018ynm,Behring:2019oci} but still, a complete procedure for 
the local subtraction of all infrared singularities, supplemented by a 
compact analytic expression for the sum of the integrated counterterms 
has been missing. In the present paper we fill this gap by providing a 
complete scheme for the pair creation of a heavy quark-antiquark pair 
out of a colourless initial state.

The paper is organized as follows. In \sect{sec:heavyQ} we 
define the problem we want to address and set up our notation. 
In \sect{sec:NLO}, we compute the NLO correction to heavy quark 
production from a general colourless initial state.
We introduce the set of NNLO counterterms in \sect{sec:NNLO}, 
where we also present a complete and compact analytic expression for 
the integrated form of the sum of all counterterms. 
In \sect{sec:Hbb}, we present the application of our subtraction 
scheme to the production of heavy quarks in the decay of the 
Standard Model Higgs boson. Finally, we draw our conclusions in 
\sect{sec:concl}.

\section{Heavy quark-antiquark pair production from colourless initial states}
\label{sec:heavyQ}

Let us consider the production of a heavy quark-antiquark pair from a generic 
colourless initial state $X$. The list of relevant subprocesses up to 
NNLO accuracy in QCD is given by

\begin{center}
\begin{tabular}{lll}
    LO & $X \to Q\bar{Q}$  & tree-level \\ & & \\
    NLO & $X \to Q\bar{Q}$  & one-loop \\
        & $X \to Q\bar{Q} g $  & tree-level \\ & & \\
    NNLO & $X \to Q\bar{Q}$  & two-loop \\
         & $X \to Q\bar{Q} g$  & one-loop \\
         & $X \to Q\bar{Q} gg$  & tree-level \\
         & $X \to Q\bar{Q} q\bar{q}$  & tree-level \\
         & $X \to Q\bar{Q} Q\bar{Q}$  & tree-level \\
\end{tabular}
\end{center}
In the above list the heavy quarks are denoted by $Q$ and $\bar{Q}$, 
while radiated gluons and light quarks are denoted by $g$ and $q\qb$ respectively. 
The matrix elements for all of these partonic processes are well-known for scalar, 
pseudo-scalar, as well as vector and axial currents. In particular the  two-loop 
form factors were first computed in \cite{Bernreuther:2004ih,Bernreuther:2004th,
Bernreuther:2005rw,Bernreuther:2005gw} up to finite terms in the parameter of 
dimensional regularization $\ep$. Recently these results have been extended up to 
$\Oe{1}$ terms in \refr{Ablinger:2017hst}.

For amplitudes, we use the colour-space notation of \cite{Catani:1996vz} 
where $\ket{}{}{}$ is an abstract vector in colour- and spin-space, such that 
the matrix element summed over colours and spins can be written as
\beq
\SME{}{}{} = \braket{}{}{}{}\,.
\eeq
Then, the insertion of colour charge operators, such as $\bT_i \bT_k$ or the 
anti-commutator $\{\bT_i \bT_k, \bT_j \bT_l\}$, as well as polarization-dependent 
tensors such as the \AP{} splitting kernels $\hP_{f_i f_r}$ (to be specified below), 
into a given amplitude interference will be denoted as
\bal
\bT_i \bT_k \otimes \SME{}{}{} &\equiv
    \bra{}{}{} \bT_i \bT_k \ket{}{}{}\,,
\\
\{\bT_i \bT_k, \bT_j \bT_l\} \otimes \SME{}{}{} &\equiv
    \bra{}{}{} \{\bT_i \bT_k, \bT_j \bT_l\} \ket{}{}{}\,,
\\
\hP_{f_i f_r} \otimes \SME{}{}{} &\equiv
    \bra{}{}{} \hP_{f_i f_r} \ket{}{}{}\,.
\eal

The amplitude $\ket{}{}{}$ has the formal loop expansion
\beq
\ket{}{}{} = \ket{}{(0)}{} + \ket{}{(1)}{} + \ket{}{(2)}{} + \ldots\,.
\eeq
We consider amplitudes computed in conventional dimensional regualrization 
with the strong coupling $\as$ and gluon wave function renormalized in the 
\MSbar{} scheme, while the heavy quark mass and wave function are renormalized 
on-shell. Furthermore, assuming $n_l$ light flavours, plus a single heavy 
flavour $Q$, we implement the transition that allows us to use $n_l+1$ active 
flavours in the running strong coupling \cite{Nason:1989zy}. 
In particular, 
\beq
\as^B \mu_0^{2\ep} = \frac{\as \mu_R^{2\ep}}{C(\ep)} 
    \left[1 - \frac{\as}{4\pi} \frac{\beta_0}{\ep} + \Oa{2}\right]\,,
\label{eq:aS-ren}
\eeq
where $\as^B$ is the bare coupling, while $\as$ denotes the renormalized 
coupling which will appear in all subsequent equations. Furthermore, 
the beta-function coefficient $\beta_0$ reads
\beq
\beta_0 = \frac{11}{3}\CA - \frac{4}{3}\TR (n_l+1)
\eeq
and
\beq
C(\ep) = (4\pi)^{\ep}\Gamma(1+\ep)\,.
\label{eq:Cep-def}
\eeq
Note that $C(\ep)$ as defined above is different form the usual 
expression in the \MSbar{} scheme of $S_\ep^{\MSbar} = (4\pi)^\ep 
\exp(-\ep \gamma_E)$ and agrees with the convention of 
\refrs{Bernreuther:2004ih,Bernreuther:2004th,Bernreuther:2005rw,Bernreuther:2005gw}. 
However, for the processes considered here the perturbative expansion starts 
at $\as^0$, therefore the inclusion of the NNLO corrections implies just one-loop 
renormalization for the coupling constant. For this reason, the $\Oe{2}$ difference 
between these two conventions turns out to have no impact on the physical result.

Throughout we denote by $P$ the total incoming momentum of the process and make use 
of the following definitions
\beq
s_{ij} \equiv 2p_i \cdot p_j\,,
\qquad
y_{ij} \equiv \frac{s_{ij}}{P^2}
\qquad\mbox{and}\qquad
y_{(ij)k} \equiv y_{ik} + y_{jk}\,.
\eeq
We also employ this notation for mapped momenta (see 
\eqns{eq:p1p2-Cirmap}{eq:4to3map} below), so that e.g.,
\beq
s_{\ha{i}\ha{k}} \equiv 2\ha{p}_i\cdot \ha{p}_k\,,
\qquad
s_{\ha{i}r} \equiv 2\ha{p}_i\cdot p_r
\qquad\mbox{and}\qquad
s_{\ti{i}\ti{k}} \equiv 2\ti{p}_i\cdot \ti{p}_k\,.
\eeq
The phase space of $n$ outgoing particles of total momentum $P$ is defined as
\beq
\PS{n}(p_1,\ldots,p_n;P) =  
    \prod_{i=1}^{n} \frac{\rd^d p_i}{(2\pi)^{d-1}} \delta_{+}(p_i^2 - m_i^2) 
    (2\pi)^d \delta^{(4)}(p_1+\ldots+p_n-P)\,.
\eeq

Finally, integrated subtractions are given in terms of multiple polylogarithms ($G$), 
which can be defined recursively by the iterated integral 
\cite{Goncharov:1998kja,Goncharov:2001iea}
\beq
G_{a_1,\ldots,a_n}(y) = 
    \int_0^y \rd t\, \frac{1}{t - a_1}G_{a_2,\ldots,a_n}(t)
\label{eq:G-def}
\eeq
with $G(y)=1$. In the special case where all the $a_i$'s are zero, 
the multiple polylogarithm is defined as
\beq
G_{0_1,\ldots,0_n}(y) = \frac{1}{n!}\ln^n y\,,
\eeq
which is consistent with  $G(y)=1$ for $n=0$. For completeness we 
note that for $a_i\in\{0,\pm 1\}$, the $G$'s are related to the 
harmonic polylogarithms ($H$) of \cite{Remiddi:1999ew} by the 
relation
\beq
G_{a_1,\ldots,a_n}(y) = (-1)^p H_{a_1,\ldots,a_n}(y)\,,
\qquad
a_i\in\{0,\pm 1\}\,
\eeq
where $p$ denotes the number of $a_i$'s that are equal to $+1$.

\section{Subtractions at NLO}
\label{sec:NLO}

Consider the NLO correction to the production of the 
heavy quark-antiquark pair, which is the sum of the real contribution 
involving the emission of an extra gluon and the virtual 
contribution containing the one-loop correction,
\beq
    \tgam{NLO}[J] = \int_3 \dgam{R} J_3 + \int_2 \dgam{V} J_2\,.
\label{eq:Gamma-NLO}
\eeq
Here $\int_3$ and $\int_2$ denote the integration over the 
$Q\bar{Q}g$ and $Q\bar{Q}$ phase space, while 
$J_3$ and $J_2$ are the values of some infrared-safe 
observable $J$ computed with the corresponding 3-~and~2-parton 
kinematics. By introducing an appropriate local subtraction term to 
regulate infrared divergences, we rewrite \eqn{eq:Gamma-NLO} 
as a sum of two finite terms  
\beq
    \tgam{NLO}[J] = \int_3 \dgam{NLO}_3 + \int_2 \dgam{NLO}_2\,,
\eeq
with the regularized real and regularized virtual 
contributions\footnote{Here and in the following, 
a regularized contribution will refer to an expression that 
is free of both explicit $\ep$-poles as well as non-integrable 
kinematic singularities.} given by
\bal
    \dgam{NLO}_3 &= \dgam{R} J_3 - \dgama{R}{1} J_2\,,
    \label{eq:dGammaNLO3}
    \\
    \dgam{NLO}_2 &= \bigg[\dgam{V} + \int_{[1]} 
    \dgama{R}{1}\bigg] 
    J_2\,.
    \label{eq:dGammaNLO2}
\eal
Above $\int_{[1]}$ denotes the integration of the subtraction 
terms over the radiation variables of the extra gluon.

\subsection{Regularized real contribution}
\label{ssec:NLO-R}

Let us consider first the real emission process, 
$X(P) \to Q(p_1) + \bar{Q}(p_2) + g(p_3)$.
Denoting by $F$ the flux factor\footnote{The flux factor
is $F=2m_X$ for the decay of a heavy particle 
$X$, while for $e^+e^-$ collisions it reads $F = 2P^2$ 
(the electron and positron are assumed to be massless).}, 
we have
\beq
    \dgam{R} = \frac{1}{F} \PS{3}(p_1,p_2,p_3;P) 
    \SME{Q\bar{Q} g}{(0)}{}
\eeq
and
\beq
    \dgama{R}{1} = \frac{1}{F} \PS{3}(p_1,p_2,p_3;P) 
    \cA_{1} \SME{Q\bar{Q} g}{(0)}{}\,.
\eeq
We note that throughout this subsection, $p_1$, $p_2$ and $p_3$ 
refer to the momenta of the heavy quark $Q$, the heavy antiquark $\bar{Q}$ 
and the gluon $g$ in the three-particle real emission phase space.
The matrix element is singular only in the $p_3^\mu \to 0$ soft gluon limit, 
thus the structure of the approximate matrix element is very simple and we 
have just one subtraction term,
\beq
    \cA_1 \SME{Q\bar{Q} g}{(0)}{} 
    = \cS{g_3}{(0)}\,.
\eeq

\paragraph{Single soft subtraction.}

The subtraction term follows the structure of the
general formula for the approximation of a tree-level $n+1$-parton 
matrix element in the $p_r^\mu \to 0$ soft limit \cite{Bassetto:1984ik} 
and 
is given by
\beq
\cS{g_r}{(0)}(p_1,p_2,p_3) \equiv 
	-8\pi\frac{\as\mu^{2\ep}}{C(\ep)}
    \sum_{\ha{i},\ha{k}} \frac{1}{2}S_{\ha{i}\ha{k}}(r)\, 
	\bT_{\ha{i}} \bT_{\ha{k}} \otimes \SME{n}{(0)}{(\ha{p}_1,\ha{p}_2)} 
	\,,
\label{eq:S3g}
\eeq
where the summation indices $\ha{i}$ and $\ha{k}$ run over the labels of 
the hard momenta that appear in the factorized matrix element (i.e., 
$\ha{1}$ and $\ha{2}$, see below), and the eikonal factor is also computed 
using these momenta,
\beq
S_{\ha{i}\ha{k}}(r) = 
	\frac{2 s_{\ha{i}\ha{k}}}{s_{\ha{i}r}s_{\ha{k}r}} = 
	\frac{(\ha{p}_i\cdot \ha{p}_k)}{(\ha{p}_i\cdot p_r)(\ha{p}_k\cdot p_r)}
\quad\mbox{and}\quad
S_{\ha{i}\ha{i}}(r) = 
	\frac{2 s_{\ha{i}\ha{i}}}{s_{\ha{i}r}^2} = 
	\frac{\ha{p}_i^2}{(\ha{p}_i\cdot p_r)^2}
	\,.
\label{eq:SiHrkHr}
\eeq

The $3\to 2$ mapping $\{p_1,p_2,p_3\} \to \{\ha{p}_1,\ha{p}_2\}$ that 
specifies the hatted momenta which enter the factorized matrix element 
in \eqn{eq:S3g} is defined as follows (to be clear, the heavy quark $Q$ 
and anitquark $\bar{Q}$ carry momenta $\ha{p}_1$ and $\ha{p}_2$ in the 
two-parton matrix element on the right hand side of \eqn{eq:S3g}),
\beq
\bsp
\ha{p}_{1}^\mu &= \Lambda^{\mu}_{\nu}(\ha{K},K) 
	\frac{1}{\Delta} (p_1^\mu - \gamma_1 p_{3}^\mu)\,,
\\
\ha{p}_{2}^\mu &= \Lambda^{\mu}_{\nu}(\ha{K},K) 
	\frac{1}{\Delta} (p_2^\mu - \gamma_2 p_{3}^\mu)\,,
\esp
\label{eq:p1p2-Cirmap}
\eeq
where $\Delta$ and $\gamma_{1,2}$ are 
\beq
\Delta = \sqrt{1 - \frac{s_{13} s_{23} s_{3P}}
	{P^2 s_{13} s_{23} - m_Q^2 s_{3P}^2}}\,,     
\eeq
and
\beq
\gamma_1 = \frac{m_Q^2 s_{23} s_{3P}} 
	{P^2 s_{13} s_{23} - m_Q^2 s_{3P}^2}\,,
	\qquad
\gamma_2 = \frac{m_Q^2 s_{13} s_{3P}} 
	{P^2 s_{13} s_{23} - m_Q^2 s_{3P}^2}\,.     
\eeq
With these definitions we have $\ha{p}_1^2=\ha{p}_2^2=m_Q^2$ and 
$\ha{K}^2 = K^2$, where
\beq
K^\mu = P^\mu
\qquad\mbox{and}\qquad
\ha{K}^\mu = 
	\frac{1}{\Delta}[p_1^\mu + p_2^\mu - (\gamma_1+\gamma_2)p_3^\mu]\,.
\eeq
Finally, $\Lambda^{\mu}_{\nu}(\ha{K},K)$ is a (proper) Lorentz transformation 
that takes $K^\mu$ into $\ha{K}^\mu$, its explicit form can be chosen e.g., as
\beq
\Lambda^{\mu}_{\nu}(K,\ha{K}) = g^{\mu}_{\nu}
	- \frac{2(K+\ha{K})^\mu (K+\ha{K})_\nu}{(K+\ha{K})^2}
	+ \frac{2 \ha{K}^\mu K_\nu}{K^2}\,.
\label{eq:Lambda}
\eeq

\subsection{Regularized virtual contribution}
\label{ssec:NLO-V}

Turning to the two terms in \eqn{eq:dGammaNLO2}, the virtual 
contribution $\dgam{V}$ involves the one-loop correction to 
the process $X(P) \to Q(p_1) + \bar{Q}(p_2)$ and we have
\beq
    \dgam{V} = \frac{1}{F} \PS{2}(p_1,p_2;P) 
    2\Re\langle{\cM}_{Q\bar{Q}}^{(0)}|\cM_{Q\bar{Q}}^{(1)}\rangle
\eeq
with
\beq
    \int_{[1]} \dgama{R}{1} = \frac{1}{F} \PS{2}(p_1,p_2;P)\, 
    \bI_{1}(p_1,p_2;\ep) \otimes \SME{Q\bar{Q}}{(0)}{}
\eeq
where the $\bI_{1}(p_1,p_2;\ep)$ operator corresponds to the integral 
of the only subtraction term, $\cS{g_3}{(0)}$. We remark that throughout this 
subsection, $p_1$ and $p_2$ denote the momenta of the heavy quark $Q$ and 
antiquark $\bar{Q}$ in the two-body phase space.

After resolving the summation in \eqn{eq:S3g} using the colour algebra
relations $\bT_1^2 = \bT_2^2 = \CF$ and $\bT_1\bT_2 = -\CF$,
the insertion operator can be computed in a straightforward way by 
integrating $\cS{g_r}{(0)}$ over the full three body phase space and 
simply divide by the volume of the (massive) two particle phase space, 
which is a constant. The result of the integration in can be cast in the form
\beq
\bI_{1}(p_1,p_2;\ep) = 
	\frac{\as}{2\pi}
	\left(\frac{\mu^2}{P^2}\right)^\ep
	\CF \left(\frac{1}{\ep} a_{-1} + a_0 + \ep\, a_1 + \Oe{2}\right)\,,
\label{eq:I10-a-coeffs}
\eeq
where the coefficients of the Laurent expansion are functions of the customary 
variable
\beq
y \equiv \frac{\sqrt{P^2} - \sqrt{P^2 - 4m_Q^2}}{\sqrt{P^2} + \sqrt{P^2 - 4m_Q^2}}\,,
\label{eq:y-def}
\eeq
which is real for the physical decay process. The coefficients appearing 
in \eqn{eq:I10-a-coeffs} above can be expressed in terms of multiple 
polylogarithms defined in \eqn{eq:G-def}, always with argument $y$. 
Setting $G_{a_1,\ldots,a_n}(y) = G_{a_1,\ldots,a_n}$ for ease of notation, 
we have
\bal
a_{-1} &=
2
+\frac{2(1+y^2)}{1-y^2} G_{0}\,,
\\
a_0 &=
   4
   -8 G_{1}
-\frac{2(1+2 y+5 y^2)}{1-y^2}G_{0}
+\frac{2(1+y^2)}{1-y^2}(4 G_{-1,0}-G_{0,0}-4 G_{0,1}+2 G_{1,0}-4 \zeta_{2})\,,
\\
a_1 &=
   8
   -16 G_{1}
   +32 G_{1,1}
-\frac{2}{1-y^2}
\Big[
   2 (1+2 y+5 y^2) G_{0}
   -8 y (1+3 y) \zeta_{2}
   +4 (3+2 y+3 y^2) G_{-1,0}
\nonumber\\& 
   -(1+2 y+5 y^2) (G_{0,0}+4 G_{0,1})
   +2 (7+2 y-y^2) G_{1,0}
   -(1+y^2) (16 G_{-1,-1,0}-4 G_{-1,0,0}
\nonumber\\& 
   -16 G_{-1,0,1}
   +8 G_{-1,1,0}-12 G_{0,-1,0}+G_{0,0,0}+4 G_{0,0,1}-14 G_{0,1,0}
   +16 G_{0,1,1}+8 G_{1,-1,0}
\nonumber\\& 
   -2 G_{1,0,0}-8 G_{1,0,1}+4 G_{1,1,0}
   -16 G_{-1} \zeta_{2}-8 G_{1} \zeta_{2}-8 \zeta_{3})
\Big]\,.
\eal
For the NLO cross section only the first two terms in the 
expansion of $\bI_{1}(p_1,p_2;\ep)$ are needed. Nevertheless, 
the order $\ep$ term will enter the integrated subtraction for 
the single unresolved limit of real-virtual emission at NNLO and 
so we present it here.

Note that in \eqn{eq:I10-a-coeffs} we have expanded the factor of 
$C(\ep)$ that appears in the denominator of \eqn{eq:S3g} which cancels, 
as usual, terms of $\gamma_E$ and $\ln(4\pi)$ coming from phase space 
integration. If the strong coupling is defined with a different 
$\ep$-dependent prefactor in \eqn{eq:aS-ren}, the explicit forms 
of the expansion coefficients change accordingly. In particular, 
adopting the standard \MSbar{} factor of 
$S_\ep^{\MSbar} = (4\pi)^\ep \exp(-\ep \gamma_E)$, we would have 
\beq
a_1^{\MSbar} = a_1 + \frac{\zeta_{2}}{2} a_{-1}\,,
\eeq
while the lower order expansion coefficients remain unchanged.

\section{Subtractions at NNLO}
\label{sec:NNLO}

The NNLO correction is the sum the double real, real-virtual 
and double virtual parts,
\beq
    \tgam{NNLO}[J] = \int_4 \dgam{RR} J_4 
    + \int_3 \dgam{RV} J_3 + \int_2 \dgam{VV} J_2\,,
\eeq
which we rearrange into three finite contributions by including 
appropriate subtraction terms,
\beq
    \tgam{NNLO}[J] = 
    \int_4 \dgam{NNLO}_4 + \int_3 \dgam{NNLO}_3
    + \int_2 \dgam{NNLO}_2\,.
\eeq
The regularized double real, regularized real-virtual and 
regularized double virtual contributions are
\bal
    \dgam{NNLO}_4 &= \dgam{RR} J_4 - \dgama{RR}{1} J_3 
    - \dgama{RR}{2} J_2 + \dgama{RR}{12} J_2
    \label{eq:dGNNLO4}
    \\
    \dgam{NNLO}_3 &= \bigg[\dgam{RV} + \int_{[1]} 
    \dgama{RR}{1}\bigg] J_3
    - \bigg[\dgama{RV}{1} 
    + \bigg(\int_{[1]} \dgama{RR}{1}\bigg)
    \strut^{\!{\rm A}_1}\bigg] J_2 
    \label{eq:dGNNLO3}\\
    \dgam{NNLO}_2 &= \bigg[\dgam{VV} + \int_{[2]} \dgama{RR}{2} 
    - \int_{[2]} \dgama{RR}{12} + \int_{[1]} \dgama{RV}{1}
    + \int_{[1]} \bigg(\int_{[1]}
    \dgama{RR}{1}\bigg)\strut^{\!{\rm A}_1}\bigg] J_2\,.
    \label{eq:dGNNLO2}
\eal
Above $\int_{[1]}$ and $\int_{[2]}$ denote the integration 
of subtraction terms over the radiation variables of one 
and two extra partons.

\subsection{Regularized double real contribution}
\label{ssec:NNLO-RR}

Considering all possible subprocesses with one heavy flavour 
$Q$ and $n_l$ massless flavours $q$, the sum of all such 
contributions reads
\beq
    \dgam{RR} = \frac{1}{F} \PS{4}(p_1,p_2,p_3,p_4;P) 
    \left[\frac{1}{2} \SME{Q\bar{Q} gg}{(0)}{}
    +n_l \SME{Q\bar{Q} q\qb}{(0)}{}
    +\frac{1}{4} \SME{Q\bar{Q} Q\bar{Q}}{(0)}{}\right]\,,
\eeq
where we have explicitly reported the statistical factors in 
front of the matrix elements for the production of two gluons 
and the one for the production of two heavy quark-antiquark pairs.
We label the particles such that the heavy quark $Q$ and heavy antiquark 
$\bar{Q}$ always carry momenta $p_1$ and $p_2$, while $p_3$ and $p_4$ are 
the momenta associated with the extra emissions (either two gluons $gg$, 
a light quark-antiquark pair $q\qb$, or one more heavy quark and 
antiquark, $Q\bar{Q}$). We emphasize that $p_1$, $p_2$, $p_3$ 
and $p_4$ denote momenta in the four-particle double real emission 
phase space throughout this subsection.
The subtraction terms introduced in \eqn{eq:dGNNLO4} 
are\footnote{Although potentially useful to stabilize 
numerical computation in the limit of small quark mass, 
in this paper we do not include subtraction terms for 
quasi-collinear limits and the production of two heavy 
quark-antiquark pairs.}
\bal
    \dgama{RR}{1} &= \frac{1}{F} \PS{4}(p_1,p_2,p_3,p_4;P) 
    \left[\frac{1}{2} \cA_1 \SME{Q\bar{Q} gg}{(0)}{}
    +n_l \cA_1 \SME{Q\bar{Q} q\qb}{(0)}{}\right]\,,
\\
    \dgama{RR}{2} &= \frac{1}{F} \PS{4}(p_1,p_2,p_3,p_4;P) 
    \left[\frac{1}{2} \cA_2 \SME{Q\bar{Q} gg}{(0)}{}
    +n_l \cA_2 \SME{Q\bar{Q} q\qb}{(0)}{}\right]\,,
\\
    \dgama{RR}{12} &= \frac{1}{F} \PS{4}(p_1,p_2,p_3,p_4;P) 
    \left[\frac{1}{2} \cA_{12} \SME{Q\bar{Q} gg}{(0)}{}
    +n_l \cA_{12} \SME{Q\bar{Q} q\qb}{(0)}{}\right]\,.
\eal

Starting with the $X(P) \to Q(p_1) + Q(p_2) + g(p_3) + g(p_4)$ subprocess, 
the $\SME{Q\bar{Q} gg}{(0)}{}$ matrix element requires regularization by 
subtraction only in the following infrared limits:
\begin{enumerate}
\item the $p_3^\mu || p_4^\mu$ single unresolved collinear limit
\item the $p_3^\mu \to 0$ soft limit
\item the $p_4^\mu \to 0$ soft limit
\item the $p_3^\mu \to 0$, $p_4^\mu \to 0$ double soft gluon limit
\end{enumerate}
Hence, the single, double and iterated single unresolved approximate matrix 
elements for this subprocess have the following structure,
\bal
\cA_{1} \SME{Q\bar{Q} gg}{(0)}{} &= 
	\cC{g_3 g_4}{(0)} \cF^{C}_{34}
	+ \left(\cS{g_3}{(0)} - \cC{g_3 g_4}{}\cS{g_3}{(0)}\right) \cF^{S}_3
	+ \left(\cS{g_4}{(0)} - \cC{g_3 g_4}{}\cS{g_4}{(0)}\right) \cF^{S}_4\,,
\label{eq:A1bbgg}
\\
\cA_{2} \SME{Q\bar{Q} gg}{(0)}{} &= \cS{g_3 g_4}{(0)}\,,
\label{eq:A2bbgg}
\\
\cA_{12} \SME{Q\bar{Q} gg}{(0)}{} &= 
	\cC{g_3 g_4}{}\cS{g_3 g_4}{(0)} \cF^{C}_{34}
	+ \left(\cS{g_3}{}\cS{g_3 g_4}{(0)} 
	    - \cC{g_3 g_4}{}\cS{g_3}{}\cS{g_3 g_4}{(0)}\right) \cF^{S}_3
	+ \left(\cS{g_4}{}\cS{g_3 g_4}{(0)} 
	    - \cC{g_3 g_4}{}\cS{g_4}{}\cS{g_3 g_4}{(0)}\right) \cF^{S}_4\,.
\label{eq:A12bbgg}
\eal
In \eqn{eq:A1bbgg}, the subtraction terms $\cC{g_3 g_4}{}\cS{g_3}{(0)}$ 
and $\cC{g_3 g_4}{}\cS{g_4}{(0)}$ are included to avoid double subtraction 
over those regions of phase space where the collinear and soft limits overlap. 
Moreover, note that formally $\cA_{12} = \cA_{1} \cA_{2}$, i.e., the 
form of the iterated single unresolved approximate matrix element 
agrees with that of the single unresolved approximate matrix element.

Although the structure of \eqnss{eq:A1bbgg}{eq:A12bbgg} is dictated by the 
types of infrared limits which require regularization, the explicit definition 
of the subtraction terms is obviously not unique. Different choices can have 
various advantages and drawbacks (e.g., locality of subtractions versus full 
analytic control over the integrated subtraction terms). In particular, a 
general issue for any subtraction scheme at NNLO concerns the integration 
of counterterms, which can turn out to be a very elaborate task. Thus, 
on practical grounds, once the general structure of the counterterms is 
defined and momentum conservation has been implemented, one may seek to 
exploit the freedom in the definitions of counterterms to simplify the 
integration. This consideration motivates the inclusion of the collinear 
factor $\cF^{C}_{34}$ and the soft factors $\cF^{S}_{3}$ and $\cF^{S}_{4}$ 
in the above formulae\footnote{Similar considerations have been 
discussed also in \refr{DelDuca:2019ctm}.}. 
Clearly, the collinear and soft factors must go 
to the identity in the corresponding limit. Furthermore, to preserve the 
structure of cancellations among the subtraction terms in all limits, 
we find that in our construction the soft-collinear overlap must be 
multipiled with the soft factor, while $\cA_{12}\SME{Q\bar{Q} gg}{(0)}{}$ 
inherits the pattern of modifications of $\cA_1\SME{Q\bar{Q} gg}{(0)}{}$.
In the following, we present a concrete example of a constructive procedure 
for obtaining factors that lead to a fully analytic result for the sum of 
all integrated subtraction terms which is very compact, see \sect{ssec:NNLO-VV}.

Turning to the $X(P) \to Q(p_1) + \bar{Q}(p_2) + q(p_3) + \qb(p_4)$ 
subprocess, the only infrared limits of $\SME{Q\bar{Q} q\qb}{(0)}{}$ 
that require regularization by subtraction are:
\begin{enumerate}
\item the $p_3^\mu || p_4^\mu$ single collinear limit
\item the $p_3^\mu \to 0$, $p_4^\mu \to 0$ double soft quark-antiquark limit
\end{enumerate}
Correspondingly, the structure of the subtractions is very simple and each 
approximate matrix element is built form a single term,
\bal
\cA_{1} \SME{Q\bar{Q} q\qb}{(0)}{} &= \cC{q_3 \qb_4}{(0)} \cF^C_{34}\,,
\label{eq:A1bbqq}
\\
\cA_{2} \SME{Q\bar{Q} q\qb}{(0)}{} &= \cS{q_3 \qb_4}{(0)}\,,
\label{eq:A2bbqq}
\\
\cA_{12} \SME{Q\bar{Q} q\qb}{(0)}{} &= \cC{q_3 \qb_4}{}\cS{q_3 \qb_4}{(0)} \cF^C_{34}\,,
\label{eq:A12bbqq}
\eal
As previously, $\cA_{12} = \cA_{1} \cA_{2}$ formally.

Before presenting the explicit expressions of each subtraction term, let us 
first discuss the kinematics and in particular the momentum mappings used to 
enforce exact phase space factorization. The definition of subtraction 
terms involves the specification of functions which map the double real emission 
phase space into phase spaces of lower multiplicity plus radiation variables. 
In particular, we find that all single unresolved subtraction terms can be 
defined using just one $4\to 3$ momentum mapping. The mapping appropriate to the 
double unresolved and iterated single unresolved subtractions is then obtained 
simply by applying the $3\to 2$ mapping discussed in \sect{ssec:NLO-R} to the 
output of the $4\to 3$ mapping presented below. Given momenta $\{p_1,p_2,p_3,p_4\}$ 
where $p_1^2=p_2^2=m_Q^2$ and $p_3^2=p_4^2=0$, to be mapped to 
$\{\ha{p}_1,\ha{p}_2,\ha{p}_{34}\}$ with $\ha{p}_1^2=\ha{p}_2^2=m_Q^2$ and 
$\ha{p}_{34}^2=0$, we set
\beq
\bsp
\ha{p}_{34}^\mu &= 1/\beta (p_3^\mu + p_4^\mu - \alpha P^\mu)\,,
\\
\ha{p}_n^\mu &= \Lambda^{\mu}_{\nu}(K,\ha{K}) p_{n}^{\nu}\,,
\qquad\qquad\qquad n=1,2\,,
\label{eq:4to3map}
\esp
\eeq
where $\alpha$ and $\beta$ are
\beq
\alpha = 
	\frac12 \left[y_{(34)P} - \sqrt{y_{(34)P}^2 - 4y_{34}}\,\right]
\qquad\mbox{and}\qquad
\beta = \frac{\sqrt{y_{(34)P}^2 - 4y_{34}}}{y_{(34)P} - y_{34}}\,.
\label{eq:al-be-def}
\eeq
With these definitions $\ha{p}_{34}$ is massless and the momenta
\beq
K^\mu = P^\mu - p_3^\mu - p_4^\mu 
\qquad\mbox{and}\qquad
\ha{K}^\mu = P^\mu - \ha{p}_{34}^\mu
\eeq
have the same mass, $K^2 = \ha{K}^2$. Hence, $\Lambda^{\mu}_{\nu}(\ha{K},K)$ 
is a (proper) Lorentz transformation that takes $K^\mu$ into $\ha{K}^\mu$, 
whose explicit form can be chosen as in \eqn{eq:Lambda}. We note that this 
mapping is equivalent to the final state mapping presented in \refr{Nagy:2007ty}.

The momentum mapping introduced above leads to the exact factorization 
of the four particle phase space in the following form,
\beq
\PS{4}(p_1,p_2,p_3,p_4;P) = 
	\PS{3}(\ha{p}_1,\ha{p}_2,\ha{p}_{34};P)[\rd p]\,,
\eeq
where the measure for the factorized radiation variables $[\rd p]$ reads
\beq
[\rd p] = x^{-1+2\ep} \frac{P^2}{2\pi} \int_{\al_{\rm min}}^{\al_{\rm max}}
	\rd \al\, (1-\al)^{-3+2\ep} (x - 2\al + \al^2)^{2-2\ep} 
	\PS{2}(p_3,p_4; \al P + \be \ha{p}_{34})\,,
\label{eq:dp}
\eeq
with 
\beq
x \equiv y_{\wha{34}P} = \frac{2\ha{p}_{34} \cdot P}{P^2}\,,
\label{eq:x-def}
\eeq
and expressing $\beta$ of \eqn{eq:al-be-def} in terms of $\al$ and 
$x$ we find
\beq
\be = \frac{x - 2\al + \al^2}{x(1-\al)}\,.
\label{eq:beta-alpha}
\eeq
Since $\ha{p}_{34}^2=0$ we have
\beq
\al_{\rm min} = 0
\qquad\mbox{and}\qquad
\al_{\rm max} = 1 - \sqrt{1-x}\,.
\label{eq:al-min-max}
\eeq

Before going on, let us anticipate some difficulties which appear when 
integrating the single unresolved subtraction terms over the measure in 
\eqn{eq:dp}. First, the definition of the collinear subtraction term 
involves the specification of a momentum fraction $z_{r}$ ($r=3,4$) 
associated with the splitting. However, a natural candidate for this variable, 
$z_{r} = \frac{p_{r}\cdot P}{(p_3+p_4)\cdot P}$,
turns out to be a somewhat complicated function of the radiation variables. 
Second, the soft subtraction term involves the eikonal factor with the hard 
momenta $\ha{p}_1$ and $\ha{p}_2$. In addition, the measure $[dp]$ evidently 
depends also on $\ha{p}_{34}$. Thus, the result of the integration will depend 
on all independent dot-products between these three vectors in a very complicated 
way. Last, the upper limit in \eqn{eq:al-min-max} is a square root function of 
the invariant $x$, which implies that the integrated counterterms will also 
be function of this square root. Regardless of the first two issues, this 
last point alone leads to difficulties when computing the iterated single 
unresolved subtraction terms.

However, exploiting the freedom in the definition of the subtraction terms, 
one can devise a strategy to tackle the above mentioned difficulties. First, 
a more convenient choice of momentum fractions can be made upon examination 
of the explicit form of the collinear integral without affecting the structure 
of singularities. Second, as anticipated in \eqn{eq:S3g} we multiply the soft 
integral by an appropriate function $\cF^S_r$ that reduces to one in the unresolved 
limit in $d$ dimensions. This function can be chosen in such a way that it cancels 
regular factors, effectively reducing the multiple angular dependence of the 
integrand. Last, one can restrict the phase space of the 
subtraction\footnote{A well-known practice in various NLO subtraction 
schemes \cite{Frixione:1995ms,Nagy:1998bb,Nagy:2003tz}.}. This restriction can be 
implemented by adopting an appropriate functional form of the upper limit of 
integration with respect to $x$ (e.g., linear) which avoids the dependence on 
square roots of invariants.

\paragraph{Single collinear subtraction.}

In order to define the collinear subtraction term, we start from the well-known 
approximation to the matrix element in this limit \cite{Altarelli:1977zs}
\beq
\SME{Q\bar{Q} f_3 f_4}{(0)}{(p_1,p_2,p_3,p_4)} \simeq 
	8\pi\frac{\as\mu^{2\ep}}{C(\ep)} \frac{1}{s_{34}} 
	\hat{P}_{f_3 f_4}(z_3, \kT{};\ep) \otimes 
	\SME{Q\bar{Q} g}{(0)}{(p_1, p_2, p_3+p_4)}
\label{eq:C34-limit}
\eeq
where $\hat{P}_{f_3 f_4}(z_3, \kT{};\ep)$ is the $d$-dimensional 
\AP{} splitting kernel for the $f_{(34)} \to f_3+f_4$ splitting 
(here $f$ denotes the parton flavour) that are functions of the momentum 
fraction ($z_3$) and the transverse momentum ($\kT{}$) of the splitting. 
For our calculation only gluon splitting is relevant, for which the kernels 
are given explicitly by
\bal
P^{\mu\nu}_{gg}(z_{3}, \kT{}) &= 
	2 \CA \bigg[ -g^{\mu\nu} \bigg(\frac{z_{3}}{1-z_{3}} + \frac{1-z_{3}}{z_{3}}\bigg) 
	- 2 (1-\ep) z_{3} (1-z_{3}) \frac{\kT{}^\mu \kT{}^\nu}{\kT{}^2}\bigg]\,,
\label{eq:Pgg}
\\
P^{\mu\nu}_{q\qb}(z_{3}, \kT{}) &= 
	\TR \bigg[ -g^{\mu\nu} 
	+ 4 z_{3} (1-z_{3}) \frac{\kT{}^\mu \kT{}^\nu}{\kT{}^2}\bigg]\,.
\label{eq:Pqq}
\eal

To build a proper subtraction counterterm from the above limit formula, as 
usual we need to evaluate the factorized matrix element on the right hand 
side with mapped momenta, that respect momentum conservation and the mass 
shell conditions. Furthermore, the momentum fractions and $\kT{}$ must be 
properly defined over the full phase space. A straightforward choice for 
$z_3$ would read
\beq
z_{3} = \frac{p_3 \cdot P}{(p_3+p_4)\cdot P}
\qquad\mbox{and}\qquad
1-z_{3} = \frac{p_4 \cdot P}{(p_3+p_4)\cdot P}\,.
\label{eq:Z34Z43}
\eeq
Although it is a simple exercise to construct the subtraction term 
in this way, using the momentum mapping of \eqn{eq:4to3map}, 
it turns out that the integrated form of this subtraction is rather 
cumbersome. In order to exhibit the reasons behind this, we recall 
the measure for the radiation variables, \eqn{eq:dp}, and note that 
the two-particle phase space $\PS{2}(p_3,p_4; \al P + \be \ha{p}_{34})$ 
appearing there can be parametrised as follows,
\beq
\PS{2}(p_3,p_4;\al P + \be \ha{p}_{34}) = \frac{1}{8\pi} 
\frac{(4\pi)^\ep}{\Gamma(1-\ep)}
	(P^2)^{-\ep} \rd v\, \al^{-\ep} (\al + \be x)^{-\ep} v^{-\ep} (1-v)^{-\ep}
	\Theta(v)\Theta(1-v)\,,
\label{eq:dphi2}
\eeq
where $\be$ is given in \eqn{eq:beta-alpha}, while $v$ is defined 
implicitly by the following relation,
\beq
z_{3} = \frac{\al(1-\al) + (x - 2\al + \al^2) v}{x - \al^2}\,.
\eeq
Notice that $1-z_{3}$ is obtained by $v \to 1-v$ in the above expression. 
Furthermore, in this parametrization the two-paricle invariant $y_{34}$ reads
\beq
y_{34} = \frac{\al(x - \al)}{1-\al}\,,
\eeq
so the colliner limit (when $y_{34}\to 0$) corresponds to $\al\to 0$ (note that 
$\al_{\rm max} = 1 - \sqrt{1-x}$ so $x-\al > 0$, since $0<x<1$), 
and in the limit $v$ is simply the momentum fraction of the splitting.

Examining the explicit forms of the \AP splitting kernels, it is clear that 
integrals involving $1/z_3$ must be evaluated, and the quadratic expression 
in the numerator of $z_3$ appears in the denominator of the integrand, 
causing the presence of square root functions of $x$ in the integrated expressions. 
Because of this, integrating these expressions further, as is necessary when computing 
the integrals of iterated single unresolved subtraction terms, becomes extremely 
complicated.

In order to avoid such complications, let us drop all terms that are not linear 
in $\al$ and in $v$ in the numerator of $z_{3}$, so that this numerator simply 
reads $\al + x v$. Enforcing the correct collinear limit ($z_3 \to v$ as $\al \to 0$) 
as well as the correspondence between $(v,z_3) \leftrightarrow (1-v,1-z_3)$, we 
define the new variable\footnote{Notice that assuming $\hz{3} = \frac{\al+ x v}{D_3}$ 
with $D_3$ independent of $v$, requiring $1-\hz{3} = \frac{\al+ x(1-v)}{D_3}$ 
immediately fixes the denominator as $D_3 = 2\al+x$.}
\beq
\hz{3} = \frac{\al + x v}{2\al + x}\,.
\label{eq:z3hat-def}
\eeq
The last requirement ensures the preservation of the symmetry between 
the daughter partons in the splitting at the integrand level. Thus our choice for 
the collinear subtraction term is
\beq
\cC{f_3 f_4}{(0)}(p_1,p_2,p_3,p_4) \equiv 
	8\pi\frac{\as\mu^{2\ep}}{C(\ep)} \frac{1}{s_{34}} 
	\hat{P}_{f_3 f_4}(\hz{3}, \hkT;\ep) \otimes 
	\SME{Q\bar{Q} g}{(0)}{(\ha{p}_1, \ha{p}_2, \ha{p}_{34})}
	\,,
\label{eq:C34-no-F34}
\eeq
where the hatted momenta appearing in the factorized matrix element 
are given in \eqn{eq:4to3map}.
We note that $\hz{3}$ can be expressed in terms of $z_3$ of \eqn{eq:Z34Z43} 
as follows,
\beq
\bsp
\hz{3} &= \frac{2 x (x + y_{34}) z_3 
	- (x - r)(x + y_{34} - r)}{2 r (2x + y_{34} - r)}\,,
\\
1-\hz{3} &= \frac{2 x (x + y_{34}) (1-z_3) 
	- (x - r)(x + y_{34} - r)}{2 r (2x + y_{34} - r)}\,,
\esp
\label{eq:hz3}
\eeq
where
\beq
r = \sqrt{(x + y_{34})^2 - 4 y_{34}}\,.
\eeq
The definition of the transverse momentum $\hkT$ that enters the \AP splitting 
kernel reads
\beq
\hkT^\mu = \zeta_{3} p_4^\mu - \zeta_{4} p_3^\mu + \zeta_{34} \ha{p}_{34}^\mu\,,
\label{eq:hkT}
\eeq
where
\beq
\zeta_{3} = z_{3} - \frac{y_{34}}{\al y_{(34)P}}\,,
\qquad
\zeta_{4} = (1-z_{3}) - \frac{y_{34}}{\al y_{(34)P}}
\qquad\mbox{and}\qquad
\zeta_{34} = \frac{y_{34}}{\al x}
	[(1 - z_3) - z_{3}]\,.
\eeq
Notice that in the above equation, we have made use of $z_3$ of \eqn{eq:Z34Z43}.
With this definition $\hkT^\mu$ is perpendicular to the parent momentum 
$\ha{p}_{34}$ and also $\hkT^\mu \to 0$ in the $p_3^\mu || p_4^\mu$ 
collinear limit.

However, even after introducing the new variable in \eqn{eq:z3hat-def}, the 
issue of square root functions of invariants in the integrated form of 
$\cC{f_3 f_4}{(0)}$ is still present due to the appearance of the factor 
$(x-2\al+\al^2)^{2-2\ep}$ in \eqn{eq:dp}. We deal with this factor by exploiting 
the freedom to multiply the subtraction term with a suitable regular function
$\cF^C_{34}$, see \eqn{eq:A1bbgg}. To make an optimal choice, we take the occasion 
to collect all factors coming from the factorized measure and the factor of $1/s_{34}$ 
which is common to all collinear integrals. Inserting the explicit expression for 
$\PS{2}(p_3,p_4;\al P + \be \ha{p}_{34})$ from \eqn{eq:dphi2} into \eqn{eq:dp}, 
we find
\beq
[\rd p]\frac{1}{s_{34}} = 
    \frac{x^{-1+2\ep}}{(4\pi)^2} \frac{(4\pi)^\ep}{\Gamma(1-\ep)} 
    (P^2)^{-\ep} \rd v\,
    \al^{-1-\ep} v^{-\ep} (1-v)^{-\ep} 
    (1-\al)^{-2+3\ep}(x-\al)^{-1-\ep} (x-2\al+\al^2)^{2-2 \ep}\,.
\eeq
We note that the product of the last three factors,
\beq
{\mathcal G}(\al,x;\ep) \equiv 
    (1-\al)^{-2+3\ep}(x-\al)^{-1-\ep} (x-2\al+\al^2)^{2-2 \ep}\,,
\eeq
does not play a role in regularizing any divergent behaviour, hence the 
integrand may be simplified (without changing the pole structure of the 
integral) by multiplying with
\beq
\frac{{\displaystyle \lim_{\al \to 0}} {\mathcal G}(\al,x;\ep)}
{{\mathcal G}(\al,x;\ep)} 
    = x^{1-3\ep}(1-\al)^{2-3\ep}(x-\al)^{1+\ep} (x-2\al+\al^2)^{-2+2 \ep}\,.
\eeq

The final source of square roots in the integral is the upper limit of 
integration in \eqn{eq:al-min-max}. Since we are free to restrict the action 
of the counterterm to a region of phase space around the singular limit, we 
choose an upper limit $\al_0(x) \le \al_{\mathrm{max}}$ such as to avoid the 
presence of square roots. One simple choice is
\beq
\al_0(x) = C\cdot\frac{x}{2} < 1 - \sqrt{1-x}\,,\qquad C,x\in (0,1]\,.
\eeq
As the final physical results cannot depend on the constant $C$, varying its 
value gives a strong check on the correct implementation of the subtraction 
scheme. Thus the final form of the regular function $\cF^C_{34}$ is given by
\beq
\cF^C_{34} \equiv x^{1-3\ep}(1-\al)^{2-3\ep}(x-\al)^{1+\ep}(x-2\al+\al^2)^{-2+2 \ep}
    \Theta[\al_0(x)-\al]\,.
\label{eq:F34-def}
\eeq

\paragraph{Single soft subtraction.}

The single soft subtraction to the double real contribution is 
structurally identical to the NLO soft subtraction term given 
in \eqn{eq:S3g} and we have,
\beq
\cS{g_r}{(0)}(p_1,p_2,p_3,p_4) \equiv 
	-8\pi\frac{\as\mu^{2\ep}}{C(\ep)}
    \sum_{\ha{i},\ha{k}} 
    \frac{1}{2}S_{\ha{i}\ha{k}}(r)\, 
	\bT_{\ha{i}} \bT_{\ha{k}} \otimes
	\SME{Q\bar{Q} g}{(0)}{(\ha{p}_1, \ha{p}_2, \ha{p}_{34})}\,.
\label{eq:S3r}
\eeq
The mapped momenta that appear in the 3-parton factorized matrix element 
above can be chosen to coincide with those used to define the collinear 
subtraction and are given in \eqn{eq:4to3map}.
We recall that the summation indices $\ha{i}$ and $\ha{k}$ in \eqn{eq:S3r} 
run over the labels of the mapped momenta that enter the factorized matrix 
element (i.e., $\ha{i},\ha{k} = \ha{1}, \ha{2}, \ha{34}$).

The integration of the soft counterterm is plagued by similar difficulties as 
the collinear case discussed above. In particular, the $(x-2\al+\al^2)^{2-2\ep}$ 
factor in \eqn{eq:dp} is present, as well as the square root in the upper limit of 
integration. As with the collinear subtraction, we can overcome these problems 
by a suitable choice of the $\cF^S_r$ function that appears in \eqn{eq:A1bbgg}. 
In order to obtain this factor, consider the most elaborate soft integral,  
which involves the eikonal factor $\frac{s_{\ha{1}\ha{2}}}{s_{\ha{1}r}s_{\ha{2}r}}$. 
It is convenient to write this integral in the rest frame of $P$, oriented such 
that $\ha{p}_{34}^\mu$ lies along the $z$-axis,
\beq
\bsp
P^\mu &= \sqrt{P^2}(1,\ldots)\,,
\\
\ha{p}_{34}^\mu &= \hat{E}_{34}(1,\ldots,1)\,,
\\
\ha{p}_1^\mu &= \hat{E}_1(1,\vec{\beta}_1)\,,
\\
\ha{p}_2^\mu &= \hat{E}_2(1,\vec{\beta}_2)\,,
\esp
\eeq
where $\ldots$ denote components that vanish\footnote{Clearly the components of 
e.g., $\ha{p}_2^\mu$ are not independent, since 
$\ha{p}_2^\mu = P^\mu - \ha{p}_1^\mu - \ha{p}_{34}^\mu$, but this will not 
play a role in what follows.}. In this frame $p_r$ reads
\beq
p_r^\mu = E_r(1,\vec{n}_r) = 
    E_r(1,\ldots\mathrm{angles}\ldots,\sin\varphi\sin\vartheta_r,
    \sin\varphi\cos\vartheta_r,\cos\vartheta_r)\,,
\eeq
where ``$\ldots\mathrm{angles}\ldots$'' are angular components on which the 
integrand does not depend. In this frame, $\PS{2}(p_3,p_4;\al P + \be \ha{p}_{34})$ 
can be written in the following form
\beq
\bsp
\PS{2}(p_3,p_4;\al P + \be \ha{p}_{34}) &= 
    2^{-3-2\ep}\pi^{-2+\ep}\frac{\Gamma(1 - \ep)}{\Gamma(1 - 2 \ep)}
	(P^2)^{-\ep} \rd \xi\, \rd\eta\, \al^{1-2\ep} \xi^{-\ep} (1-\xi)^{-\ep}
	\eta^{-\frac12-\ep}(1-\eta)^{-\frac12-\ep}
\\ &\times
	(\al + \be x)^{1-2\ep} (\al+\be x \xi)^{-2+2\ep} 
	\Theta(\xi)\Theta(1-\xi)\Theta(\eta)\Theta(1-\eta)\,,
\label{eq:dphi2-2}
\esp
\eeq
where
\beq
\cos\vartheta_r=1-2\xi
\qquad\mbox{and}\qquad
\cos\varphi=1-2\eta\,.
\eeq
Then we find
\beq
\frac{s_{\ha{1}\ha{2}}}{s_{\ha{1}3}s_{\ha{2}3}} = 
    \frac{2\ha{p}_1\cdot \ha{p}_2}{(2\ha{p}_1\cdot p_r)(2\ha{p}_1\cdot p_r)} =
    \frac{2\ha{p}_1\cdot \ha{p}_2}{[2\hat{E}_1 E_r (1-\vec{\beta}_1\cdot\vec{n}_r)]
    [2\hat{E}_2 E_r (1-\vec{\beta}_2\cdot\vec{n}_r)]}
\eeq
while the energy $E_r$ takes the form
\beq
E_r = \frac{\al(\al+\be x)}{\al+\be x \xi} \sqrt{P^2}\,.
\eeq
Hence (using $\hat{E}_1 = y_{\ha{1}P} P^2/2$ and $\hat{E}_2 = y_{\ha{2}P} P^2/2$)
\beq
\bsp
[\rd p] \frac{s_{\ha{1}\ha{2}}}{s_{\ha{1}3}s_{\ha{2}3}} &= 
    2^{-4 - 2 \ep} \pi^{-3 + \ep} \frac{\Gamma(1 - \ep)}{\Gamma(1 - 2 \ep)} 
    (P^2)^{-\ep} \frac{y_{\ha{1}\ha{2}}}{y_{\ha{1}P} y_{\ha{2}P}} 
    \rd\xi\, \rd\eta\,
    \al^{-1 - 2 \ep} \frac{\xi^{-\ep} (1 - \xi)^{-\ep}  \eta^{-\frac12 - \ep} 
    (1 - \eta)^{-\frac12 - \ep}}{(1-\vec{\beta}_1\cdot\vec{n}_r)
    (1-\vec{\beta}_2\cdot\vec{n}_r)} 
\\ &\times
    x^{-1 + 2 \ep} (1 - \al)^{-2 + 2 \ep} (x - \al)^{-1 - 2 \ep} 
    (x -2 \al + \al^2)^{2 - 2 \ep} [\al(1-\al) + (x - 2 \al + \al^2)\xi]^{2 \ep}\,.
\esp
\eeq
We note that the product of factors on the second line,
\beq
{\mathcal G}(\al,x,\xi;\ep) \equiv 
    x^{-1 + 2 \ep} (1 - \al)^{-2 + 2 \ep} (x - \al)^{-1 - 2 \ep} 
    (x -2 \al + \al^2)^{2 - 2 \ep} [\al(1-\al) + (x - 2 \al + \al^2)\xi]^{2 \ep}\,,
\eeq
does not play a role in regularizing any divergent behaviour, hence the 
integrand may be simplified (without altering its pole structure) if we 
multiply it with
\beq
\bsp
\frac{{\displaystyle \lim_{\al \to 0}} {\mathcal G}(\al,x,\xi;\ep)}
{{\mathcal G}(\al,x,\xi;\ep)} 
    &= \xi^{2\ep} x^{1-2\ep} (1-\al)^{2-2\ep} 
	(x - \al)^{1+2\ep} (x - 2\al + \al^2)^{-2+2\ep}  
\\ &\times
	[\al (1-\al) + (x - 2\al + \al^2) \xi]^{-2\ep}\,.  
\esp
\eeq

As was the case with the collinear subtraction term, the upper limit of 
integration again leads to the appearance of square roots in the integral. 
Following the same strategy as in the case of the collinear subtraction, 
we arrive at the following formula for $\cF^S_r$\,
\beq
\bsp
\cF^S_{r} &\equiv \xi^{2\ep} x^{1-2\ep} (1-\al)^{2-2\ep} 
	(x - \al)^{1+2\ep} (x - 2\al + \al^2)^{-2+2\ep}  
\\ &\times
	[\al (1-\al) + (x - 2\al + \al^2) \xi]^{-2\ep}
    \Theta[\al_0(x)-\al]\,.
\esp
\label{eq:Fr-def}
\eeq
With the above choice of $\cF^S_r$, the soft integral can be performed 
to yield a fully analytic and reasonably compact expression which is 
suitable for further integration, as is necessary when computing the 
integrated forms of the iterated single unresolved counterterms.

\paragraph{Single soft-collinear overlap.} 

The only single unresolved subtraction term in \eqn{eq:A1bbgg} 
that we have not yet specified is the soft-collinear overlap 
$\cC{g_3 g_4}{}\cS{g_r}{(0)}$. Our choice is
\bal
\cC{g_3 g_4}{}\cS{g_3}{(0)}(p_1,p_2,p_3,p_4) \equiv 
	8\pi\frac{\as\mu^{2\ep}}{C(\ep)} \frac{2}{s_{3\ha{34}}}
	\frac{1-z_{3,\ha{34}}}{z_{3,\ha{34}}}\, \CA\,
	 \SME{Q\bar{Q} g}{(0)}{(\ha{p}_1, \ha{p}_2, \ha{p}_{34})}\,,
\label{eq:C34S3}
\\
\cC{g_3 g_4}{}\cS{g_4}{(0)}(p_1,p_2,p_3,p_4) \equiv 
	8\pi\frac{\as\mu^{2\ep}}{C(\ep)} \frac{2}{s_{4\ha{34}}}
	\frac{1-z_{4,\ha{34}}}{z_{4,\ha{34}}}\, \CA\,
	 \SME{Q\bar{Q} g}{(0)}{(\ha{p}_1, \ha{p}_2, \ha{p}_{34})}\,.
\label{eq:C34S4}
\eal
Note that in this subtraction term, the momentum fractions must be evaluated 
with hatted momenta, so that they match the soft subtraction in the 
collinear limit. Hence the momentum fractions $z_{3,\ha{34}}$ and 
$z_{4,\ha{34}}$ are defined as
\beq
z_{3,\ha{34}} = \frac{p_3 \cdot P}{(\ha{p}_{34} + p_3) \cdot P}
\qquad\mbox{and}\qquad
z_{4,\ha{34}} = \frac{p_4 \cdot P}{(\ha{p}_{34} + p_4) \cdot P}\,.
\label{eq:z334z434}
\eeq
The mapped momenta entering the factorized matrix elements in 
\eqns{eq:C34S3}{eq:C34S4} are once again given by 
\eqn{eq:4to3map}.

We can now clarify the reason that the soft-collinear overlap terms in 
\eqn{eq:A1bbgg} have to be multiplied with the same $\cF^{S}_r$ functions 
as the soft subtractions. In the $p_r^\mu \to 0$ soft limit both 
$\cF^{C}_{34} \to 1$ as well as $\cF^{S}_r \to 1$. 
Thus $\cC{g_3 g_4}{}\cS{g_r}{(0)}$ properly regularizes $\cC{g_3 g_4}{(0)}$ 
in the soft limit. On the other hand, in the $p_3^\mu || p_4^\mu$ collinear 
limit $\cF^{S}_r \not\to 1$ in $d$ dimensions. So to insure the proper 
cancellation of $\cS{g_r}{(0)}$ with $\cC{g_3 g_4}{}\cS{g_r}{(0)}$ in the 
collinear limit, the latter must be multiplied by the same factor of $\cF^{S}_r$ 
as the former.

\paragraph{Double soft subtraction.}

Turning to the double unresolved subtraction, we recall that only the double 
soft limit requires regularization by subtraction. We choose to define the 
subtraction term for this limit as follows. For double soft gluon emission 
we define
\beq
\bsp
\cS{g_3 g_4}{(0)}(p_1,p_2,p_3,p_4) &\equiv 
	\left[8\pi\frac{\as\mu^{2\ep}}{C(\ep)}\right]^2
	\bigg\{
	\sum_{i,j,k,l=1,2}
	\frac{1}{8}S_{ik}(3) S_{jl}(4)\,
	\{\bT_{\ti{i}} \bT_{\ti{k}} , \bT_{\ti{j}} \bT_{\ti{l}}\}
\\
	&- \frac14 \CA 
	\sum_{i,k=1,2}
	\Big[S_{ik}(3,4) - S_{ik}^{\rm mass}(3,4)\Big] \bT_{\ti{i}} \bT_{\ti{k}}
	\bigg\} \otimes
	\SME{Q\bar{Q} }{(0)}{(\ti{p}_1,\ti{p}_2)}\,,
\label{eq:S3g4g}
\esp
\eeq
while for a soft quark-antiquark pair we set
\beq
\bsp
\cS{q_3 \qb_4}{(0)}(p_1,p_2,p_3,p_4) &\equiv 
	\left[8\pi\frac{\as\mu^{2\ep}}{C(\ep)}\right]^2 
	\frac{1}{s_{34}^2}
	\TR 
\\ &\times
	\sum_{i,k=1,2}
	\frac{s_{i3} s_{k4} + s_{i4} s_{k3} - s_{ik} s_{34}}
	{(s_{i3}+s_{i4}+s_{34})(s_{k3}+s_{k4}+s_{34})}
	\bT_{\ti{i}} \bT_{\ti{k}} \otimes
	\SME{Q\bar{Q} }{(0)}{(\ti{p}_1,\ti{p}_2)}\,.
\label{eq:S3q4q}
\esp
\eeq
The eikonal factors $S_{ik}(r)$ and $S_{jl}(r)$ read
\beq
S_{ik}(3) = \frac{2s_{ik}}{s_{i3} s_{k3}} 
	= \frac{(p_i\cdot p_k)}{(p_i\cdot p_3)(p_k\cdot p_3)}
\qquad\mbox{and}\qquad
S_{jl}(4) = \frac{2s_{jl}}{s_{j4} s_{l4}} 
	= \frac{(p_j\cdot p_l)}{(p_j\cdot p_4)(p_l\cdot p_4)}\,.
\eeq
Furthermore, for $S_{ik}(3,4)$ we have
\beq
S_{ik}(3,4) = 
	S_{ik}^{\rm (s.o.)}(3,4) 
	+ 4\frac{s_{i3} s_{k4} + s_{i4} s_{k3}}
		{(s_{i3} + s_{i4} + s_{34})(s_{k3} + s_{k4} + s_{34})}
	\bigg[\frac{1-\ep}{s_{34}^2} - \frac18 S_{ik}^{\rm (s.o.)}(3,4) \bigg]
	- \frac{4}{s_{34}} S_{ik}(34)\,,
\label{eq:Sikrs}
\eeq
where the $S_{ik}^{\rm (s.o.)}(3,4)$ is the strongly-ordered limit of this 
expression in either the $p_3^\mu\to 0$ or $p_4^\mu\to 0$ limit (it is symmetric 
in $3$ and $4$),
\beq
S_{ik}^{\rm (s.o.)}(3,4) =
    S_{ik}(4)\Big(S_{i4}(3) + S_{k4}(3) - S_{ik}(3)\Big) =
	\frac{4s_{ik}}{s_{i3} s_{k4} s_{34}} + \frac{4s_{ik}}{s_{i4} s_{k3} s_{34}}
	- \frac{4s_{ik}^2}{s_{i3} s_{i4} s_{k3} s_{k4}}
\eeq
and
\beq
S_{ik}(34) = \frac{2s_{ik}}
	{(s_{i3} + s_{i4} + s_{34})(s_{k3} + s_{k4} + s_{34})}\,.
\eeq
Last, $S_{ik}^{\rm mass}(3,4)$ is directly proportional to the square of the 
heavy quark mass,
\beq
\bsp
S_{ik}^{\rm mass}(3,4) &= 
	\frac{s_{i3}}{s_{i3} + s_{i4} + s_{34}} 
	S_{ik}^{\rm mass, (s.o.)}(3,4) 
	+ \frac{s_{k4}}{s_{k3} + s_{k4} + s_{34}} 
	S_{ki}^{\rm mass, (s.o.)}(4,3) 
\\ &
	-\frac{2}{s_{34}} 
	\frac{s_{i3} s_{k4} + s_{i4} s_{k3}}
		{(s_{i3} + s_{i4} + s_{34})(s_{k3} + s_{k4} + s_{34})}
	\bigg[\frac{s_{ii}}{s_{i3} s_{i4}} + \frac{s_{kk}}{s_{k3} s_{k4}}\bigg]\,,
\esp
\eeq
where $S_{ik}^{\rm mass, (s.o.)}(3,4)$ and 
$S_{ki}^{\rm mass, (s.o.)}(4,3)$ are the strongly-ordered limits of 
$S_{ik}^{\rm mass}(3,4) $ in the $p_4^\mu \to 0$ and $p_3^\mu \to 0$ limit,
\bal
S_{ik}^{\rm mass, (s.o.)}(3,4) &= 
	S_{ii}(3) \Big(S_{k3}(4) - S_{ik}(4)\Big)\,,
\\
S_{ki}^{\rm mass, (s.o.)}(4,3) &= 
	S_{kk}(4) \Big(S_{i4}(3) - S_{ik}(3)\Big)\,.
\eal
Summations over all indices in \eqn{eq:S3g4g} run over 
$i,j,k,l=1,2$ and the equivalence of any and all indices is allowed.

We remark that contrary to the choice in \eqns{eq:S3g}{eq:S3r}, here the 
hard momenta $p_i$, $p_k$, $p_j$ and $p_l$ that appear in the various functions 
just defined are simply the original momenta of the heavy quarks in the 
four-particle phase space and not the mapped momenta. This choice is quite 
convenient for the present calculation, since it allows us to use known results 
for massive four particle phase space integrals \cite{Bernreuther:2011jt,
Bernreuther:2013uma} to compute the integrated subtraction term. For similar 
reasons, we prefer to define the subtractions in \eqns{eq:S3g4g}{eq:S3q4q} by 
retaining the subleading (in the double soft limit) $s_{34}$ term in 
denominators of the form $(s_{i3}+s_{i4}+s_{34})$ and $(s_{k3}+s_{k4}+s_{34})$ 
throughout. Thus, our subtraction terms differ by these subleading terms from the 
double soft limit formulae of \cite{Catani:1999ss,Czakon:2014oma}.

To complete the definition of the subtraction term, we must specify the 
momenta $\ti{p}_1$ and $\ti{p}_2$ that enter the factorized matrix element. 
Starting from the four momenta of the double real emission phase space, 
we apply the $4 \to 3$ mapping of \eqn{eq:4to3map}, followed 
by the $3 \to 2$ mapping presented in \eqn{eq:p1p2-Cirmap} in order to obtain 
$\ti{p}_1$ and $\ti{p}_2$.

Finally, as remarked above, all master integrals that are needed to compute 
the integrated double soft subtraction term are known in the literature 
\cite{Bernreuther:2011jt,Bernreuther:2013uma}, and so we find it most 
convenient to not include any additional factors with the double soft 
subtraction, see \eqns{eq:A2bbgg}{eq:A2bbqq}.

\paragraph{Single collinear--double soft subtraction.}

In order to cancel the singularities of the double soft subtraction term 
in the single collinear limit, as well as the singularities of the single 
collienar subtraction term in the double soft limit, we introduce the iterated 
single unresolved counterterm
\beq
\bsp
\cC{f_3 f_4}{}\cS{f_4 f_4}{(0)}(p_1,p_2,p_3,p_4) &\equiv
    \left[8\pi\frac{\as\mu^{2\ep}}{C(\ep)}\right]^2
	\frac{1}{s_{34}}
	\sum_{\ha{i},\ha{k} = \ha{1},\ha{2}}
	\frac{2 \ha{p}_{i,\mu} \ha{p}_{k,\nu}}{s_{\ha{i}\ha{34}} s_{\ha{k}\ha{34}}}
\\ &\times
	P_{f_3 f_4}^{\mu\nu}(\hz{3},\hkT{};\ep) \bT_{\ti{i}} \bT_{\ti{k}} \otimes
	\SME{Q\bar{Q}}{(0)}{(\ti{p}_1,\ti{p}_2)}\,,
\esp
\label{eq:Cf3f4Sf3f4}
\eeq
where the \AP kernels are given in \eqns{eq:Pgg}{eq:Pqq}, while $\hz{3}$ 
and $\hkT^\mu$ are defined in \eqns{eq:hz3}{eq:hkT}. The momenta $\ha{p}_i$, 
$\ha{p}_k$ and $\ha{p}_{34}$ that appear in the uncontracted eikonal factor 
above are those obtained by the $4\to 3$ mapping of 
\eqn{eq:4to3map}, while the factorized matrix element 
is evaluated with the same momenta $\ti{p}_1$ and $\ti{p}_2$ that enter 
the definition of the double soft subtraction term.

We remark that this term  enters \eqn{eq:A12bbgg} multiplied with the 
factor of $\cF^{C}_{34}$. Since this function goes to one in the 
collinear limit, $\cC{q_r \qb_s}{}\cS{q_r \qb_s}{(0)}$ correctly 
regularizes $\cS{q_r \qb_s}{(0)}$ in this limit. On the other hand, 
in the double soft limit $\cF^{C}_{34} \not\to 1$, 
so $\cC{g_r g_s}{}\cS{g_r g_s}{(0)}$ must be multiplied with $\cF^{C}_{34}$ 
to ensure the proper cancellation of this term with 
$\cC{g_r g_s}{(0)}\cF^{C}_{34}$ in the double soft limit.

\paragraph{Single soft--double soft subtraction.}

The iterated single soft--double soft subtraction term regularizes the 
double soft subtraction term in the $p^\mu_s \to 0$ single soft limit, 
as well as the single soft subtraction term $\cS{g_s}{(0)}$ in the 
double soft limit,
\beq
\bsp
\cS{g_s}{}\cS{g_r g_s}{(0)}(p_1,p_2,p_3,p_4) &\equiv 
    \left[8\pi\frac{\as\mu^{2\ep}}{C(\ep)}\right]^2
	\bigg\{
	\sum_{\ha{i},\ha{k},\ha{j},\ha{l} = \ha{1},\ha{2}}
	\frac{1}{8}S_{\ha{i}\ha{k}}(\ha{34}) S_{\ha{j}\ha{l}}(s)\,
	\{\bT_{\ti{i}} \bT_{\ti{k}} , \bT_{\ti{j}} \bT_{\ti{l}}\}
\\ &
	- \frac14 \CA \sum_{\ha{i},\ha{k} = \ha{1},\ha{2}} 
	\Big[S_{\ha{i}\ha{k}}(\ha{34})\Big(S_{\ha{i}\ha{34}}(s) 
		+ S_{\ha{k}\ha{34}}(s) - S_{\ha{i}\ha{k}}(s)\Big)
\\ & 
		- S_{\ha{i}\ha{i}}(\ha{34})\Big(S_{\ha{k}\ha{34}}(s) 
		- S_{\ha{i}\ha{k}}(s)\Big)
	\Big] \bT_{\ti{i}} \bT_{\ti{k}}
	\bigg\} \otimes
	\SME{Q\bar{Q} }{(0)}{(\ti{p}_1,\ti{p}_2)}\,.
\esp
\label{eq:SsgSrgsg}
\eeq
As always, summation indices can be equal. For the sake of clarity, we emphasize that
\beq
S_{\ha{i}\ha{k}}(\ha{34}) = 
	\frac{(\ha{p}_i\cdot \ha{p}_k)}{(\ha{p}_i\cdot \ha{p}_{34})
	(\ha{p}_k\cdot \ha{p}_{34})}
\qquad\mbox{and}\qquad
S_{\ha{i}\ha{i}}(\ha{34}) = 
	\frac{\ha{p}_i^2}{(\ha{p}_i\cdot \ha{p}_{34})^2}\,.
\eeq
Furthermore we have e.g.,
\beq
S_{\ha{j}\ha{l}}(s) = 
	\frac{(\ha{p}_j\cdot \ha{p}_l)}{(\ha{p}_j\cdot p_s)(\ha{p}_l\cdot p_s)}
\qquad\mbox{and}\qquad
S_{\ha{i}\ha{34}}(s) = 
	\frac{(\ha{p}_i\cdot \ha{p}_{34})}{(\ha{p}_i\cdot p_s)(\ha{p}_{34}\cdot p_s)}\,,
\eeq
with obvious generalizations for the other terms not displayed explicitly. 
Here the set of hatted momenta are obtained from the original momenta via 
the $4\to 3$ mapping given in \eqn{eq:4to3map}. The 
tilded momenta entering the factorized matrix element are again equal to 
those in the double soft subtraction.

Let us remark that this term enters \eqn{eq:A12bbgg} multiplied with 
a factor of $\cF^{S}_s$. Since this function goes to one as $p_s^\mu \to 0$, 
$\cS{g_s}{}\cS{g_r g_s}{(0)}$ regularizes $\cS{g_r g_s}{(0)}$ correctly in 
this limit. On the other hand, in the double soft limit, $\cF^S_{s} \not\to 1$, 
hence $\cS{g_s}{}\cS{g_r g_s}{(0)}$ must be multiplied by the same factor 
as $\cS{g_s}{(0)}$ in order to ensure the cancellation of these terms 
in the double soft limit.

\paragraph{Soft-collinear--double soft overlap.}

The set of subtractions listed so far leads to double subtraction 
in the soft-collinear limit. In order to avoid this, we introduce 
the last counterterm in \eqn{eq:A12bbgg}, given by
\bal
\cC{g_3 g_4}{}\cS{g_3}{}\cS{g_3 g_4}{(0)}(p_1,p_2,p_3,p_4) &\equiv 
	-\left[8\pi\frac{\as\mu^{2\ep}}{C(\ep)}\right]^2
	\sum_{\ha{i},\ha{k} = \ha{1},\ha{2}} 
	\frac12 S_{\ha{i}\ha{k}}(\ha{34}) 
\\ &\times
	\frac{2}{s_{3\ha{34}}} 
	\frac{1-z_{3,\ha{34}}}{z_{3,\ha{34}}}\,
	\CA\, \bT_{\ti{i}} \bT_{\ti{k}} \otimes
	\SME{Q\bar{Q} }{(0)}{(\ti{p}_1,\ti{p}_2)}\,,
\label{eq:C34S3gS3g4g}
\\
\cC{g_3 g_4}{}\cS{g_4}{}\cS{g_3 g_4}{(0)}(p_1,p_2,p_3,p_4) &\equiv 
	-\left[8\pi\frac{\as\mu^{2\ep}}{C(\ep)}\right]^2
	\sum_{\ha{i},\ha{k} = \ha{1},\ha{2}} 
	\frac12 S_{\ha{i}\ha{k}}(\ha{34})
\\ &\times
	\frac{2}{s_{4\ha{34}}} 
	\frac{1-z_{4,\ha{34}}}{z_{4,\ha{34}}}\,
	\CA\, \bT_{\ti{i}} \bT_{\ti{k}} \otimes
	\SME{Q\bar{Q} }{(0)}{(\ti{p}_1,\ti{p}_2)}\,.
\label{eq:C34S4gS3g4g}
\eal
Above $z_{3,\ha{34}}$ and $z_{4,\ha{34}}$ are defined in \eqn{eq:z334z434}. 
As before, hatted momenta are obtained from the $4\to 3$ mapping of 
\eqn{eq:4to3map}, while the tilded momenta are given  
as in the double soft subtraction term, by applying the $3 \to 2$ mapping of 
\eqn{eq:p1p2-Cirmap}.

The factor multiplying $\cC{g_r g_s}{}\cS{g_s}{}\cS{g_r g_s}{(0)}$ in 
\eqn{eq:A12bbgg} is doubly constrained. In fact, this term 
must regularize $\cS{g_s}{}\cS{g_r g_s}{(0)}\cF^{S}_s$ in the collinear 
limit as well as $\cC{g_r g_s}{}\cS{g_s}{(0)}\cF^{S}_{s}$ in the double 
soft limit. Moreover $\cF^{S}_s\not\to 1$ in either of these limits, 
so we must multiply $\cC{g_r g_s}{}\cS{g_s}{}\cS{g_r g_s}{(0)}$ by $\cF^{S}_s$ 
in \eqn{eq:A12bbgg} to achieve the correct pattern of cancellations.

\paragraph{Pattern of cancellations.} 

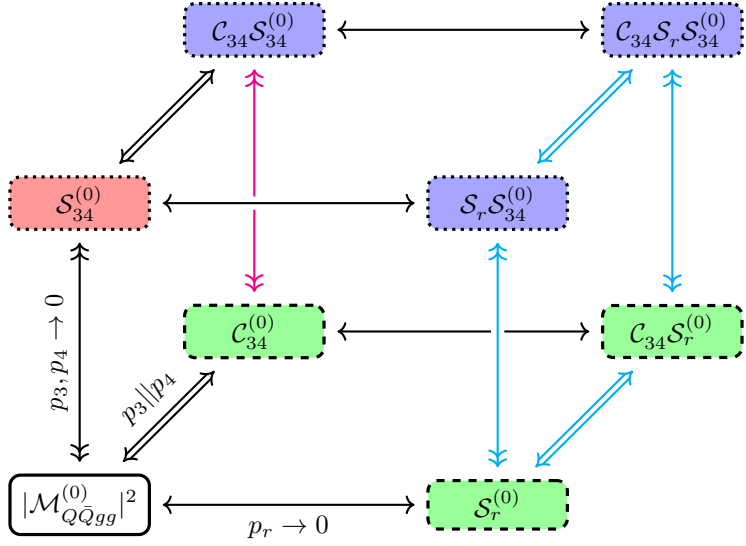
\begin{figure}
    \centering
\begin{tikzpicture}

\node[align=center, draw=black, rounded corners, fill=white, outer sep=0.5em, 
text width = 1.6cm, very thick] at (0,0) (me) {$\SME{Q\bar{Q} gg}{(0)}{}$};
\node[align=center, draw=black, rounded corners, fill=white, outer sep=0.5em, 
text width = 1.6cm, dotted, very thick, fill=red!40] at (0,4) (s34) {$\cS{34}{(0)}$};
\node[align=center, draw=black, rounded corners, fill=white, outer sep=0.5em, 
text width = 1.6cm, dashed, very thick, fill=green!40] at (5.5,0) (sr) {$\cS{r}{(0)}$};
\node[align=center, draw=black, rounded corners, fill=white, outer sep=0.5em, 
text width = 1.6cm, very thick, dotted, fill=blue!35] at (5.5,4) (srs34) {$\cS{r}{}\cS{34}{(0)}$};

\draw (me)+(45:3.25) node[align=center, draw=black, rounded corners, 
fill=white, outer sep=0.5em, text width = 1.6cm, very thick, dashed, fill=green!40]
(c34) {$\cC{34}{(0)}$};

\draw (sr)+(45:3.25) node[align=center, draw=black, rounded corners, 
fill=white, outer sep=0.5em, text width = 1.6cm, very thick, dashed,fill=green!40] 
(c34sr) {$\cC{34}{}\cS{r}{(0)}$};

\draw (s34)+(45:3.25) node[align=center, draw=black, rounded corners, 
fill=white, outer sep=0.5em, text width = 1.6cm, very thick, dotted, fill=blue!35] 
(c34s34) {$\cC{34}{}\cS{34}{(0)}$};

\draw (srs34)+(45:3.25) node[align=center, draw=black, rounded corners, 
fill=white, outer sep=0.5em, text width = 1.6cm, very thick, dotted, fill=blue!35] 
(c34srs34) {$\cC{34}{}\cS{r}{}\cS{34}{(0)}$};

\draw[double distance=1.5pt, Implies-Implies,thick] (me) -- (c34) node[midway] 
(coll) {};
\draw[<->,thick] (me) -- (sr) node[midway] (soft) {};
\draw[double distance=1.5pt, Implies-Implies,thick,cyan] (sr) -- (c34sr);
\draw[<->,thick] (c34) -- (c34sr);

\draw[<<->>,thick] (me) -- (s34) node[midway] (doublesoft) {};
\draw[white,line width=5pt] (c34) -- (c34s34);
\draw[<<->>,thick,magenta] (c34) -- (c34s34);
\draw[white,line width=5pt] (sr) -- (srs34);
\draw[<<->>,thick,cyan] (sr) -- (srs34);
\draw[<<->>,thick,cyan] (c34sr) -- (c34srs34);

\draw[white,line width=5pt] (s34) -- (c34s34);
\draw[double distance=1.5pt, Implies-Implies,thick] (s34) -- (c34s34);
\draw[white,line width=5pt] (s34) -- (srs34);
\draw[<->,thick] (s34) -- (srs34);
\draw[double distance=1.5pt, Implies-Implies,thick,cyan] (srs34) -- (c34srs34);
\draw[<->,thick] (c34s34) -- (c34srs34);

\draw (coll)+(135:0.4) node[align=center,rotate=45] {$p_3||p_4$};
\draw (soft)+(0,-0.3) node[align=center] {$p_r \to 0$};
\draw (doublesoft)+(-0.3,0) node[align=center,rotate=90] {$p_3,p_4\to 0$};
    
\end{tikzpicture}

\caption{Schematic view of the pattern of cancellations 
among the matrix element and subtraction terms for double 
real emission. 
\label{fig:RR-line}
}
\end{figure}

Finally, the pattern of cancellations in the various limits 
among the matrix element and subtraction terms we have introduced 
is schematically illustrated in \fig{fig:RR-line} for the case 
of double gluon emission. The three directions identify the three 
singular limits, namely single soft (horizontal arrows), 
single collinear (diagonal double arrows) and double soft 
(vertical arrows with double arrowheads). The green, red 
and blue boxes represent $\cA_1$, $\cA_2$ and $\cA_{12}$-type 
subtraction terms. The contour of the boxes reflects the multiplicity 
of the phase space over which the observable is computed: solid, 
dashed or dotted for $J_4$, $J_3$ and $J_2$ respectively. 
Last, the magenta vertical arrow connects 
the two subtraction terms that are multiplied by $\cF^{C}_{34}$, 
while the cyan arrows connect counterterms that are multiplied 
by $\cF^{S}_{r}$. Since for light quark-antiquark pair emission 
only the single colliear and double soft limits require regularization, 
in that case only the four leftmost terms in \fig{fig:RR-line} are 
present.

\subsection{Regularized real-virtual contribution}
\label{ssec:NNLO-RV}

The real-virtual contribution to the differential decay rate involves 
one-loop corrections to the process $X(P) \to Q(p_1) + \bar{Q}(p_2) + g(p_3)$ 
and takes the form
\beq
    \dgam{RV} = \frac{1}{F} \PS{3}(p_1,p_2,p_3;P) 
    2\Re\braket{Q\bar{Q} g}{(0)}{(1)}{}\,,
\eeq
while the three other terms appearing in \eqn{eq:dGNNLO3} can be written as
\bal
    \int_{[1]} \dgama{RR}{1} &= 
        \frac{1}{F} \PS{3}(p_1,p_2,p_3;P) \bI_{1}(p_1,p_2,p_3;\ep) \otimes 
        \SME{Q\bar{Q} g}{(0)}{}\,,
\\
    \dgama{RV}{1} &= \frac{1}{F} \PS{3}(p_1,p_2,p_3;P) 
        \cA_1 \Big[2\Re\braket{Q\bar{Q} g}{(0)}{(1)}{}\Big]\,, 
\\
    \bigg(\int_{[1]} \dgama{RR}{1}\bigg)\strut^{\!{\rm A}_1} &=
        \frac{1}{F} \PS{3}(p_1,p_2,p_3;P) \cA_1\Big[\bI_{1}(p_1,p_2,p_3;\ep) \otimes 
        \SME{Q\bar{Q} g}{(0)}{}\Big]\,.
\label{eq:A1-IxM0}        
\eal
Let us emphasize that throughout this subsection, $p_1$, $p_2$ and $p_3$ denote 
the momenta of the heavy quark, heavy antiquark and gluon in the three-particle 
real emission phase space.

Starting with $\int_{[1]} \dgama{RR}{1}$, we note that due to the presence of 
the factors $\cF^C_{34}$ and $\cF^S_{r}$, the integrated counterterms can 
be computed straightforwardly by a direct evaluation of their corresponding parametric 
integral representations. To perform the integrations and manipulate the output, we 
used the {\tt PolyLogTools} package of \refr{Duhr:2019tlz}. In the case of the soft 
subtraction, the angular integrals that appeared were evaluated using the 
results of \refr{Somogyi:2011ir}. After gathering all contributions, we find 
that the insertion operator can be written as
\beq
\bsp
\bI_{1}(p_1,p_2,p_3;\ep) = 
    \frac{\as}{2\pi}
	\left(\frac{\mu^2}{P^2}\right)^\ep    
	\bigg[&
    \CA \IcC{g}{}(y_{3P};\ep) 
    +(\CA - 2\CF) \IcS{mm}{(1,2)}(y_{1P},y_{2P},w;\ep)
\\&
    -\CA \Big(
         \IcS{m0}{(1,3)}(y_{13},y_{3P};\ep)
        +\IcS{m0}{(2,3)}(y_{23},y_{3P};\ep)
    \Big)
\\&
    + \CF \Big(
         \IcS{m}{(1,1)}(y_{13},y_{3P};\ep)
        +\IcS{m}{(2,2)}(y_{23},y_{3P};\ep)
    \Big)
    \bigg]\,,
\esp
\label{eq:I10-RR}
\eeq
where the variable $w$ is defined by
\beq
w = \sqrt{1-\frac{4m_Q^2}{s_{12}+2m_Q^2}}\,.
\eeq
In order to write \eqn{eq:I10-RR}, we used the colour algebra 
relations $\bT_1^2=\bT_2^2 = \CF$, $\bT_3^2=\CA$ together with 
\beq
\bT_1 \bT_2 = \frac{\CA-2\CF}{2}
\qquad\mbox{and}\qquad
\bT_1 \bT_3 = \bT_2 \bT_3 = -\frac{\CA}{2}\,.
\eeq
The various functions that enter \eqn{eq:I10-RR} are as follows. First, the 
integrated single collinear and single soft-collinear subtraction terms are 
assembled into the function $\IcC{g}{}$,
\beq
\IcC{g}{}(x;\ep) = 
    \frac{1}{2} [\IcC{34}{}]_{gg}(x;\ep)
    +n_l [\IcC{34}{}]_{q\qb}(x;\ep)
    -[\IcC{34}{}\IcS{r}{}](x;\ep)
\eeq
with
\bal
[\IcC{34}{}]_{gg}(x;\ep) &=
    \frac{2}{\ep^2}
    + \bigg[\frac{11}{3} - 4 \ln(x)\bigg]\frac{1}{\ep}
    + \frac{271}{36} - \frac{25}{6}  \ln{2} + 12 \ln{3} 
\nonumber\\&
    - \frac{22}{3} \ln(x) + 4 \ln^2(x) + 4 \Li_2(-2) - 4 \zeta_{2}
    + \Oe{1}\,,
\\
[\IcC{34}{}]_{q\qb}(x;\ep) &=
    -\frac{2}{3\ep}
    - \frac{43}{36} - \frac{5}{6}  \ln{2} + \frac{4}{3} \ln(x)
    + \Oe{1}\,,
\\
[\IcC{34}{}\IcS{r}{}](x;\ep) &=
    -\frac{1}{\ep^2}  
    + \frac{2}{\ep} \ln\left(\frac{x}{2}\right) 
    - 2 \ln^2\left(\frac{x}{2}\right)
    + \Oe{1}\,.
\eal
The single soft subtraction involves a double summation over hard  
momenta, so the integrated soft counterterm also takes the form of a sum, 
where the contributions correspond to the integrated eikonal function 
involving two different massive hard momenta, $\IcS{mm}{(1,2)}$, one 
massive and one massless hard momentum $\IcS{m0}{(i,r)}$ and finally 
a single massive hard momentum, $\IcS{m}{(i,i)}$,
\bal
&
\IcS{mm}{(1,2)}(x_{1},x_{2},w;\ep) =
    -\frac{1 + w^2}{2w}\ln\left(\frac{1 - w}{1 + w}\right)\frac{1}{\ep}
    +     \frac{\left(1+w^2\right)}{8 w}
\nonumber\\ &\quad\times
 \bigg\{
    4\Li_2\left(\frac{(1-w)\left(x_2(1-x_2)-x_1(1-x_1)\right)}
        {x_1 (w(2-x_1-x_2)+x_1-x_2)}\right)
    -2\Li_2\left(\frac{(1-w) ((1+w)x_2-(1-w)x_1)}
        {4 w x_1}\right)
\nonumber\\ &\quad\quad
    -4\Li_2\left(\frac{(1+w)\left(x_1(1-x_1)-x_2(1-x_2)\right)}
        {x_1 (w(2-x_1-x_2)+x_2-x_1)}\right)
    +2\Li_2\left(\frac{(1+w)((1+w) x_1-(1-w) x_2)}
        {4 w x_1}\right)
\nonumber\\ &\quad\quad
    + 2\ln \left(\frac{(1-w) x_1}{(1+w) x_2}\right)
        \ln\left(\frac{(1+w)x_2-(1-w)x_1}{(w(2-x_1-x_2)+x_2-x_1)^2}\right)
    +\ln^2(1-w)
    -\ln ^2(1+w)
\nonumber\\ &\quad\quad
    +\ln\left(\frac{1-w}{1+w}\right) 
        \bigg[2 \ln (w (x_1+x_2-1))
        +8 \ln (2-x_1-x_2)
        -\ln (16 x_1 x_2)
    \bigg]
    + (1 \leftrightarrow 2)
    \bigg\} + \Oe{1}\,,
\\
&
\IcS{m0}{(i,r)}(y_{ir},y_{rP};\ep) =
    \frac{1}{2\ep^2} 
    + \frac{1}{2} \ln\left(\frac{4 P^2 y_{ir}^2}{m_Q^2 y_{rP}^4}\right)\frac{1}{\ep}
    + \frac{1}{4} \ln^2\left(\frac{4 P^2 y_{ir}^2}{m_Q^2 y_{rP}^4}\right) 
    + \Oe{1}\,,
\\
&
\IcS{m}{(i,i)}(y_{ir},y_{rP};\ep) = 
    2 \ep\,\IcS{m0}{(i,r)}(y_{ir},y_{rP};\ep)\,.
\eal
With these definitions, it is straightforward to show that the sum
\beq
\dgam{RV} + \int_{[1]} \dgama{RR}{1}
\label{eq:RV+RRA1}
\eeq
is free of explicit poles in $\ep$. However, both terms develop 
non-integrable singular behaviour as the gluon becomes soft. We 
deal with these divergences by introducing appropriate subtraction 
terms.
Since only the $p^\mu_3\to 0$ soft gluon limit requires regularization, 
the structure of the approximate matrix elements are very simple,
\bal
    \cA_1 \Big[2\Re\braket{Q\bar{Q} g}{(0)}{(1)}{}\Big] 
    &= \cS{g_3}{(1)}\,,
\\
    \cA_{[1]}\Big[\bI_{1}(p_1,p_2,p_3;\ep) \otimes 
        \SME{Q\bar{Q} g}{(0)}{}\Big]
    &= \cS{g_3}{(I\otimes 0)}\,.
\eal

\paragraph{Real-virtual single soft subtraction.}

Starting with the real-virtual contribution, we consider the general 
expression for the soft current at one-loop for massive amplitudes 
computed in \cite{Bierenbaum:2011gg,Czakon:2018iev}. Collecting terms 
in this formula that do not automatically vanish by colour conservation 
and using the $3 \to 2$ momentum mapping $\{p_1,p_2,p_3\} \to \{\ha{p}_1,\ha{p}_2\}$ 
of \eqn{eq:p1p2-Cirmap}, our choice for the counterterm is given by
\beq
\bsp
\cS{g_r}{(1)}(p_1,p_2,p_3) &\equiv 
	-8\pi\frac{\as\mu^{2\ep}}{C(\ep)}
	\bigg\{ \sum_{\ha{i},\ha{k}=\ha{1},\ha{2}}
	\frac{1}{2}S_{\ha{i}\ha{k}}(r)\, \bT_{\ha{i}} \bT_{\ha{k}} \otimes 
	2\Re\braket{Q\bar{Q}}{(0)}{(1)}{(\ha{p}_1,\ha{p}_2)}
\\&
	+2\CA
	\sum_{\substack{\ha{i},\ha{k}=\ha{1},\ha{2} \\ \ha{i} \ne \ha{k}}} 
	\left[\frac{1}{2}S_{\ha{i}\ha{k}}(r) - \frac{1}{2}S_{\ha{i}\ha{i}}(r)\right] 
	R_{\ha{i}\ha{k}} \bT_{\ha{i}} \bT_{\ha{k}} \otimes
	\SME{Q\bar{Q}}{(0)}{(\ha{p}_1,\ha{p}_2)}
\\&
	 - \frac{\as}{2 \pi} \frac{1}{C(\ep)} \frac{1}{2 \ep}
    \bigg[\beta_0 
    + \frac{4}{3} \TR \bigg(\frac{\mu_R^2}{m_Q^2}\bigg)^\ep\bigg]
	\sum_{\ha{i},\ha{k}=\ha{1},\ha{2}}
	\frac{1}{2}S_{\ha{i}\ha{k}}(r)\, \bT_{\ha{i}} \bT_{\ha{k}} \otimes 
	\SME{Q\bar{Q}}{(0)}{(\ha{p}_1,\ha{p}_2)}
	\bigg\}
	\,,
\esp
\label{eq:S3g-1-loop}
\eeq
where the definition of the eikonal factor in \eqn{eq:SiHrkHr} is recalled 
here for convenience,
\beq
S_{\ha{i}\ha{k}}(r) = 
	\frac{2 s_{\ha{i}\ha{k}}}{s_{\ha{i}r}s_{\ha{k}r}} = 
	\frac{(\ha{p}_i\cdot \ha{p}_k)}{(\ha{p}_i\cdot p_r)(\ha{p}_k\cdot p_r)}
\qquad\mbox{and}\qquad
S_{\ha{i}\ha{i}}(r) = 
	\frac{2 s_{\ha{i}\ha{i}}}{s_{\ha{i}r}^2} = 
	\frac{\ha{p}_i^2}{(\ha{p}_i\cdot p_r)^2}\,.
\eeq
Note that the contribution on the last line of \eqn{eq:S3g-1-loop} contains 
the terms that arise form the renormalization of the one-loop soft current.
The one-loop function $R_{\ha{i}\ha{k}}$ can be written in the following form,
\beq
R_{\ha{i}\ha{k}} = \frac{\as}{2\pi} 
	\left(\frac{1}{2}S_{\ha{i}\ha{k}}(r) \mu^2\right)^\ep
	\left[-\frac{1}{2\ep^2} -\frac{1}{2}\sum_{n=-1}^{1} \ep^n 
	R_{\ha{i}\ha{k}}^{(n)}\right]\,,
\label{eq:R_ik-hat}
\eeq
where we have adapted the prefactor to our conventions. The functional forms of 
the $R_{\ha{i}\ha{k}}^{(n)}$ coefficients are taken from \cite{Czakon:2018iev}, 
\allowdisplaybreaks
\bal
R_{\ha{i}\ha{k}}^{(-1)} &= 
\lvph{} - \frac{\vmh}{\ha{v}} \Big(\lnih{} + \lnjh{}\Big) \,,
\\
R_{\ha{i}\ha{k}}^{(0)} &= 
\frac{1}{\ha{v}} \Bigg[ \frac{1}{\dijh} \Big( (\aih \vph - \ajh
\vmh) \lnih2 + \big( \ajh \vph - \aih \vmh \big) \lnjh2 \Big) 
\\&
+ \Big( \lnih{} + \lnjh{} \Big) \big( \vph \lvph{} - \lvh \big) - \lixh2
\Bigg] + \frac{1}{2} \lvph2 + \zeta_{2} \Big( \frac{7}{\ha{v}} - \frac{37}{4}
\Big) \,, 
\nonumber\\
R_{\ha{i}\ha{k}}^{(1)} &= 
\frac{1}{\dijh} \Bigg\{ \Big(1 - \dijh \Big) \Bigg[
\lniph{} \lnih2 + \lnjph{} \lnjh2 
\nonumber\\&
+  2 \Big( \lnih{} \liih2 + \lnjh{} \lijh2 \Big) - \lixh2 \Big( \lnih{} +
\lnjh{} \Big) 
\nonumber\\&
+  2 \Big( \lixh3 - \liih3 - \lijh3 + \zeta_{3} \Big) \Bigg] - 7 \zeta_{2}
\Big( \lnih{} + \lnjh{} \Big) 
\nonumber\\&
+ \frac{1}{\ha{v}} \Bigg[ \Big( \big(\ajh \vph - \aih \vmh \big) \lnih2 + \big(
\aih \vph - \ajh \vmh \big) \lnjh2 \Big) \lvph{} 
\nonumber\\&+ 
\big( \aih - \ajh \big)
\Big( \lnih2 - \lnjh2 \Big) \lvh - \Big( \dih \lnih{} + \djh \lnjh{} \Big)
\big( \lixh2 
\nonumber\\&
- 7 \zeta_{2} \big) \Bigg] \Bigg\}
+ \frac{1}{\ha{v}} \Bigg\{ \Bigg[ \lvph{} \Big( \frac{3 + \ha{v}}{4} \lvph{} - \lvh
\Big) - 4 \vmh \zeta_{2} \Bigg] \Big( \lnih{} + \lnjh{} \Big) 
\nonumber\\&
-\frac{\vmh}{6} \Big( \lnih3 + \lnjh3 \Big) + 2 \lixph3 + \lixh3 
+ 12 \zeta_{2} \lvh 
\nonumber\\&
- \Bigg[ \lixh2 +
\zeta_{2} \Big( 5 + 10 \ha{v} \Big) \Bigg] \lvph{}
\Bigg\} + \frac{1}{6} \lvph3 - \Big( \frac{5}{2} +
\frac{1}{\ha{v}} \Big) \zeta_{3} \,.
\eal
Due to the different choice of prefactors, the forms given 
above are not exactly equal to those in \refr{Czakon:2018iev}. In 
particular, $R_{\ha{i}\ha{k}}^{(1)}$ differs from the expression 
presented in \cite{Czakon:2018iev} by terms proportional to 
$\zeta_{2}R_{\ha{i}\ha{k}}^{(-1)}$. However, we note that the order 
$ep$ coefficient $R_{\ha{i}\ha{k}}^{(1)}$ is only relevant to compute 
the integrated subtraction term, but otherwise does not enter 
the regularized real-virtual contribution that is actually integrated 
numerically in four dimensions. The variables in the equations 
above are defined as \cite{Czakon:2018iev}
\beq
\begin{gathered}
\aih \equiv \frac{m_i^2  \, s_{\ha{k}r}}{s_{\ha{i}\ha{k}} s_{\ha{i}r}} \,, \quad
\ajh \equiv \frac{m_k^2 \, s_{\ha{i}r}}{s_{\ha{i}\ha{k}} s_{\ha{k}r}} \,, \quad
\dih \equiv 1 - 2 \aih \,, \quad \djh \equiv 1 - 2 \ajh \,, 
\\
\ha{v} \equiv \sqrt{1 - 4 \aih \ajh} \; , \quad \ha{v}_{\pm} \equiv \frac{1 \pm
  \ha{v}}{2} \,, \quad  \ha{x} \equiv \frac{\vmh}{\vph} \,.
\end{gathered}
\label{eq:1-loop-vars-hat}
\eeq
Notice, that similarly to the tree-level single soft subtraction term in 
\eqn{eq:S3g}, the eikonal factors and variables are computed using the hard 
momenta which appear in the factorized matrix elements in \eqn{eq:S3g-1-loop}.
In our specific case, this leads to simplifications, since $\ha{v}$ 
reduces to a function of just the (fixed) heavy quark mass and $P^2$.
In particular, we find
\beq
\ha{v} = \frac{1 - y^2}{1 + y^2}
\eeq
in terms of the variable $y$ of \eqn{eq:y-def}.

\paragraph{Single soft subtraction to the integrated single unresolved 
counterterm.}

Finally, the subtraction for the integrated single unresolved 
counterterm, \eqn{eq:A1-IxM0} can be easily defined for our 
purpose as follows:
\beq
\cS{g}{(I\otimes 0)}(p_1,p_2,p_3) = 
    -8\pi\frac{\as\mu^{2\ep}}{C(\ep)}
    \sum_{\ha{i},\ha{k}=\ha{1},\ha{2}} \frac{1}{2}S_{\ha{i}\ha{k}}(r)\, 
	\bT_{\ha{i}} \bT_{\ha{k}} \otimes 
	\bI_{1,S}(p_1,p_2,p_3;\ep) \SME{Q\bar{Q}}{(0)}{(\ha{p}_1,\ha{p}_2)}\,.
\label{eq:S3g-Ix0}
\eeq
Similarly to the real-virtual single soft subtraction term, the momenta 
entering the factorized matrix element in \eqn{eq:S3g-Ix0} above are obtained 
from the $3\to 2$ momentum mapping of \eqn{eq:p1p2-Cirmap}.
In the soft limit, the insertion operator reads
\beq
\bsp
\bI_{1,S}(p_1,p_2,p_3;\ep) = 
    \frac{\as}{2\pi}
	\left(\frac{\mu^2}{P^2}\right)^\ep    
	\bigg[&
    \CA \IcC{g}{}(y_{3P};\ep) 
    +(\CA - 2\CF) \IcS{mm}{(1,2)}(y_{1P},y_{2P},w_S;\ep)
\\&
    -\CA \Big(
         \IcS{m0}{(1,3)}(y_{13},y_{3P};\ep)
        +\IcS{m0}{(2,3)}(y_{23},y_{3P};\ep)
    \Big)
\\&
    + \CF \Big(
         \IcS{m}{(1,1)}(y_{13},y_{3P};\ep)
        +\IcS{m}{(2,2)}(y_{23},y_{3P};\ep)
    \Big)
    \bigg]\,,
\esp
\eeq
and the only difference between $\bI_{1,S}$ and $\bI_{1}$  
is that $\IcS{mm}{(1,2)}$ must be evaluated with the variable $w$ computed 
in the soft limit, that we denote as $w_S$. Expressing $w_S$ with the variable 
$y$ of \eqn{eq:y-def}, we find
\beq
w_S = \frac{1 - y}{1 + y}\,.
\eeq

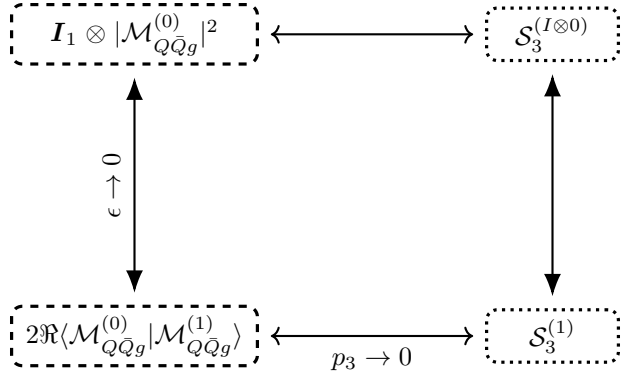
\begin{figure}
    \centering
\begin{tikzpicture}

\node[align=center, draw=black, rounded corners, fill=white, outer sep=0.5em, 
text width = 3cm, dashed, very thick] at (0,0) (me) {$2\Re\braket{Q\bar{Q} g}{(0)}{(1)}{}$};
\node[align=center, draw=black, rounded corners, fill=white, outer sep=0.5em, 
text width = 3cm, dashed, very thick] at (0,4) (s34) {$\bI_{1} \otimes \SME{Q\bar{Q} g}{(0)}{}$};
\node[align=center, draw=black, rounded corners, fill=white, outer sep=0.5em, 
text width = 1.5cm, dotted, very thick] at (5.5,0) (sr) {$\cS{3}{(1)}$};
\node[align=center, draw=black, rounded corners, fill=white, outer sep=0.5em, 
text width = 1.5cm, dotted, very thick] at (5.5,4) (srs34) {$\cS{3}{(I\otimes 0)}$};

\draw[<->,thick] (me) -- (sr) node[midway] (soft) {};

\draw[{Latex[length=3mm]}-{Latex[length=3mm]},thick] (me) -- (s34) 
node[midway] (doublesoft) {};
\draw[{Latex[length=3mm]}-{Latex[length=3mm]},thick] (sr) -- (srs34);
\draw[<->,thick] (s34) -- (srs34);
\draw (soft)+(0,-0.3) node[align=center] {$p_3 \to 0$};
\draw (doublesoft)+(-0.3,0) node[align=center,rotate=90] {$\ep \to 0$};

\end{tikzpicture}

\caption{Schematic view of the cancellations among the 
real-virtual matrix element, the integrated single unresolved 
subtraction term, as well as the corresponding soft counterterms. 
}
\label{fig:RV-line}
\end{figure}

\paragraph{Pattern of cancellations.} 

To finish this subsection, we illustrate the structure of cancellations 
among the various terms in \fig{fig:RV-line}. The soft limit of the 
real-virtual matrix element is regulated in $d$ dimensions by the 
$\cS{3}{(1)}$ subtraction defined in \eqn{eq:S3g-1-loop}. We note that 
the difference $\left[2\Re\braket{Q\bar{Q} g}{(0)}{(1)}{} - \cS{3}{(1)}\right]$, 
although free of non-integrable kinematical singularities is not finite 
in $\ep$. Similarly, $\bI_{1} \otimes \SME{Q\bar{Q} g}{(0)}{}$ is 
regularized in the soft limit by the subtraction term $\cS{g}{(I\otimes 0)}$ 
in \eqn{eq:S3g-Ix0}, but the difference 
$\left[\bI_{1} \otimes \SME{Q\bar{Q} g}{(0)}{} - \cS{g}{(I\otimes 0)}\right]$ 
still contains poles in $\ep$. However, since the $\ep$ poles of 
$2\Re\braket{Q\bar{Q} g}{(0)}{(1)}{}$ and 
$\bI_{1} \otimes \SME{Q\bar{Q} g}{(0)}{}$ explicitly cancel 
(see \eqn{eq:RV+RRA1}), we must have that the explicit poles in 
$\left[\cS{3}{(1)} + \cS{g}{(I\otimes 0)}\right]$ also cancel. This can 
be easily verified directly, using the explicit expressions presented 
above. In \fig{fig:RV-line}, the cancellation of explicit $\ep$-poles 
is represented by vertical arrows with full arrowheads, while the 
regularization of kinematic singularities in the single soft limit 
is denoted by horizontal arrows. The contour of the boxes again 
reflects the multiplicity of the phase space over which the observable 
is computed: dashed or dotted for $J_3$ and $J_2$ respectively.

\subsection{Regularized double virtual contribution}
\label{ssec:NNLO-VV}

Finally, the regularized double virtual contribution is the sum of 
the two-loop corrections to the $X(P)\to Q(p_1) + \bar{Q}(p_2)$ process 
and the four integrated counterterms that we have not yet discussed, 
see \eqn{eq:dGNNLO2}. Note that throughout this subsection, $p_1$ 
and $p_2$ denote the momenta of the heavy quark $Q$ and the heavy 
antiquark $\bar{Q}$.

In order to integrate the remaining subtraction terms, we follow a dual 
strategy. First, the double soft subtraction terms can be reduced to known 
four-particle massive phase space integrals \cite{Bernreuther:2011jt,
Bernreuther:2013uma} via integration-by-parts (IBP) identities. The IBP 
reduction is rather straightforward and so the integrated double soft 
subtraction can be obtained easily. As for the rest of the necessary 
integrated subtraction terms, we performed a direct evaluation of their 
various parametric representations, similarly to the case of the single 
unresolved subtraction terms discussed in \ref{ssec:NNLO-RV}.

The collection of all integrated counterterms in \eqn{eq:dGNNLO2} can 
be written in the following form
\beq
\bsp
&
\int_2 \dgama{RR}{2} 
    - \int_2 \dgama{RR}{12} + \int_1 \dgama{RV}{1}
    + \int_1 \left(\int_1 \dgama{RR}{1}\right)\strut^{\!{\rm A}_1} = 
\\
&\qquad\qquad=
\bigg\{\bI_{2}(p_1,p_2;\ep) - 
    \frac{\as}{2 \pi} \frac{1}{C(\ep)}\frac{1}{2 \ep} 
    \bigg[\beta_0
        + \frac{4}{3} \TR \bigg(\frac{\mu_R^2}{m_Q^2}\bigg)^\ep\bigg]
    \bI_{1}(p_1,p_2;\ep) \bigg\}
    \otimes \SME{Q\bar{Q}}{(0)}{}
\\&\qquad\qquad
+ \bI_{1}(p_1,p_2;\ep)
    \otimes 2\Re\braket{Q\bar{Q}}{(0)}{(1)}{}\,.
\esp
\label{eq:IVV-sum}
\eeq
Notice that the term proportional to 
$\bI_{1}(p_1,p_2;\ep) \otimes \SME{Q\bar{Q}}{(0)}{}$ on the second line 
corresponds precisely to the renormalization counteterm of the one-loop 
soft current in \eqn{eq:S3g-1-loop}. We find it convenient to keep this 
term explicit for an easy conversion to the case of multiple heavy quarks.

The $\bI_{1}(p_1,p_2;\ep)$ insertion operator has been given explicitly 
in \eqn{eq:I10-a-coeffs}, while $\bI_{2}(p_1,p_2;\ep)$ takes the following 
form,
\bal
\bI_{2}(p_1,p_2;\ep) = \left[\frac{\as}{2 \pi}
    \left(\frac{\mu_R^2}{P^2}\right)^\ep\right]^2
    \bigg\{&
    \CF^2 \left(\frac{b_{-2}}{\ep^2} + \frac{b_{-1}}{\ep} + b_0\right)
    +
    \CF \CA \left(\frac{c_{-2}}{\ep^2} + \frac{c_{-1}}{\ep} + c_0\right)
\\&+
    \CF \TR n_l \left(\frac{d_{-2}}{\ep^2} + \frac{d_{-1}}{\ep} + d_0\right)
    + \Oe{1}
    \bigg\}\,,
\label{eq:I20}
\eal
where the coefficients of the Laurent expansion are functions of the $y$ variable 
given in \eqn{eq:y-def}. The $b$, $c$ and $d$ coefficients are as follows 
(as before we use compact notation $G_{a_1,\ldots,a_n} = G_{a_1,\ldots,a_n}(y)$),
\allowdisplaybreaks
\bal
b_{-2} &=
    2
+\frac{4(1+y^2)}{1-y^2}G_{0}
+\frac{4(1+y^2)^2}{(1-y^2)^2}G_{0,0}
\\
b_{-1} &=
   8
   -16 G_{1}
+\frac{4(1-3 y)}{1-y}G_{0}
+\frac{8(1+y^2)}{1-y^2}(2 G_{-1,0}-4 G_{0,1}-G_{1,0}-2 \zeta_{2})
-\frac{4(1+y^2)}{(1-y^2)^2}
\nonumber\\&
\Big[
   (3+4 y+9 y^2) G_{0,0}
   -(1+y^2) (8 G_{-1,0,0}+4 G_{0,-1,0}-3 G_{0,0,0}-8 G_{0,0,1}
   -2G_{0,1,0}
\nonumber\\&   
   +4 G_{1,0,0}-4 G_{0} \zeta_{2})
\Big]
\\
b_0 &=
   -156
   -64 (G_{1}+\ln{2})
   +128 G_{1,1}
-\frac{32(1-3 y)}{1-y}G_{0,1}
-\frac{4}{1-y^2}
\Big[
   2 (11+20 y+19 y^2) G_{0}
\nonumber\\&
   +(11-23 y-25 y^2) \zeta_{2}
   +12 (1+y^2) G_{0} \ln{2}
   +2 (19-3 y+19 y^2) G_{-1,0}
   +2 (17+2 y+y^2) G_{1,0}
\nonumber\\&   
   +4 (1+y^2) (8 G_{-1,0,1}-2 G_{-1,1,0}-16 G_{0,1,1}-2 G_{1,-1,0}
   -4G_{1,0,1}-7 G_{1,1,0}+2 G_{1} \zeta_{2}
\nonumber\\&   
   +2 G_{-1,0} \ln{2}+2 G_{1,0} \ln{2}
   -\zeta_{2} \ln{2})
   +4 (8-y+8 y^2) G_{-1} \zeta_{2}
   -8 (2+y+2 y^2) G_{-1,-1,0}
\Big]
\nonumber\\&   
-\frac{8(1-9 y+17 y^2-25 y^3)}{(1-y) (1-y^2)}G_{1,0,0}
-\frac{8(23+27 y) (1+y^2)}{(1+y) (1-y^2)}G_{0,1,0}
-\frac{4}{5 (1-y^2)^2}
\nonumber\\&
\Big[
   5 (16-y-20 y^2-19 y^3+64 y^4) G_{0,0}
   -20 (2+3 y+4 y^2+5 y^3+6 y^4) \zeta_{3}
   +40 (1+3 y^4) G_{0,0} \ln{2}
\nonumber\\&
   -10 (1+10 y+24 y^2+12 y^3+17 y^4) G_{0} \zeta_{2}
   +20 (9+3 y+24 y^2+y^3+11 y^4) G_{-1,0,0}
\nonumber\\&   
   +40 (8+5 y+10 y^2+4 y^3+5 y^4) G_{0,-1,0}
   -5 (3+8 y+44 y^2-4 y^3+13 y^4) G_{0,0,0}
\nonumber\\&   
   -40 (1+y^2) (3+4 y+9 y^2) G_{0,0,1}
   -(1+y^2) (101+132 y^2) \zeta_{2}^2
   -10 (1+y^2)^2 (40 G_{-1,-1,0,0}
\nonumber\\&   
   +28 G_{-1,0,-1,0}-32 G_{-1,0,0,1}
   -8 G_{-1,0,1,0}+10 G_{0,-1,-1,0}-16 G_{0,-1,0,1}+4 G_{0,-1,1,0}
\nonumber\\&
+12 G_{0,0,0,1}-23 G_{0,0,1,0}+32 G_{0,0,1,1}+4 G_{0,1,-1,0}
+8 G_{0,1,0,1}+14 G_{0,1,1,0}+24 G_{1,0,-1,0}
\nonumber\\&
+6 G_{1,0,0,0}-16 G_{1,0,0,1}+4 G_{1,0,1,0}-24 G_{1,1,0,0}
-10 G_{-1,0} \zeta_{2}-15 G_{0,-1} \zeta_{2}-4 G_{0,1} \zeta_{2}
\nonumber\\&
-8 G_{1,0} \zeta_{2}-4 G_{-1} \zeta_{3}-12 G_{1} \zeta_{3}
+8 G_{-1,0,0} \ln{2}-4 G_{0,-1,0} \ln{2}-4 G_{0,1,0} \ln{2}+8 G_{1,0,0} \ln{2}
\nonumber\\&
+2 G_{0} \zeta_{2} \ln{2}+4 \zeta_{3} \ln{2})
   -10 (1+y^2) (1+5 y^2) G_{0} \zeta_{3}
   -10 (2-y^2) (1+y^2) G_{0,0} \zeta_{2}
\nonumber\\&
   +80 y^2 (1+y^2) G_{0,0,0} \ln{2}
   +20 (1+y^2) (3+11 y^2) G_{-1,0,0,0}
   +10 (1+y^2) (21+17 y^2) G_{0,-1,0,0}
\nonumber\\&
   +60 (1+y^2) (5+6 y^2) G_{0,0,-1,0}
   -5 (1+y^2) (7+9 y^2) G_{0,0,0,0}
   +10 (9-7 y^2) (1+y^2) G_{0,1,0,0}
\Big]
\\
c_{-2} &=
\frac{11}{6}
+\frac{11 (1+y^2)}{6 (1-y^2)}G_{0}
\\
c_{-1} &=
    \frac{181}{18}
   -\frac{44}{3} G_{1}
+\frac{1}{18 (1-y^2)}
\Big[
   (1-132 y-263 y^2) G_{0}
   -6 (47+71 y^2) \zeta_{2}
\nonumber\\&   
   +12 (1+y^2) (19 G_{-1,0}
   -22 G_{0,1}+8 G_{1,0})
   -6 (11-y^2) G_{0,0}
\Big]
\nonumber\\&   
+\frac{(1+y^2)}{(1-y^2)^2}
\Big[
   (1+y^2) (-2 G_{0,-1,0}-2 G_{0,1,0}+\zeta_{3})
   -(1+9 y^2) G_{0} \zeta_{2}
   +4 y^2 G_{0,0,0}
\Big]
\\
c_0 &=
\frac{3277}{27}
-\frac{724}{9}G_{1}
+\frac{121}{3} \ln{2}
- 24\ln{3}
+\frac{352}{3}G_{1,1}
+\frac{1}{27 (1-y^2)}
\Big[
   (541+642 y-1631 y^2) G_{0}
\nonumber\\&      
   -216 (1-y^2+G_{0}+y^2 G_{0}) (\Li_{2}(-2) + \zeta_{2})
  -3 (370-471 y-746 y^2) \zeta_{2}
  -648 (1+y^2) G_{0} \ln{3}
\nonumber\\&        
  +9 (73+48 y+73 y^2) G_{0} \ln{2}
  -6 (88+381 y+88 y^2) G_{-1,0}
  -12 (1-132 y-263 y^2) G_{0,1}
\nonumber\\&
  -12 (97+81 y-167 y^2) G_{1,0}
  -108 (16-y+16 y^2) G_{-1} \zeta_{2}
  -72 (7-17 y^2) G_{1} \zeta_{2}
\nonumber\\&
  -36 (1+y^2) (76 G_{-1,0,1}-12 G_{-1,1,0}-88 G_{0,1,1}-12G_{1,-1,0}
  +32 G_{1,0,1}+9 G_{1,1,0}
\nonumber\\&
  -24 G_{-1,0} \ln{2}-12 G_{1,0} \ln{2})
  +72 (49+3 y+49 y^2) G_{-1,-1,0}
  +72 (11-y^2) G_{0,0,1}
\nonumber\\&
  +648 (1+y^2) G_{0,-1,0,0}
\Big]
+\frac{1}{90 (1-y^2)^2}
\Big[
   10 (77-99 y-534 y^2-657 y^3+565 y^4) G_{0,0}
\nonumber\\& 
   -60 (163+54 y+12 y^2+42 y^3-55 y^4) \zeta_{3}
   -120 (2+12 y-18 y^2-15 y^3-107 y^4) G_{0} \zeta_{2}
\nonumber\\&
   +720 (1+7 y^4) G_{0,0} \ln{2}
   -120 (35+45 y+39 y^3+y^4) G_{-1,0,0}
   -240 (17-12 y-30 y^2-15 y^3
\nonumber\\&   
   -38 y^4) G_{0,-1,0}
   +60 (23-6 y-48 y^2+48 y^3+91 y^4) G_{0,0,0}
   -60 (91-60 y-72 y^2-60 y^3
\nonumber\\&   
   -91 y^4) G_{0,1,0}
   -60 (115+72 y+48 y^2+72 y^3+173 y^4) G_{1,0,0}
   -9 (109+136 y-54 y^2-136 y^3
\nonumber\\&   
   -163 y^4) \zeta_{2}^2
   -180 (1+y^2)^2 (16 G_{-1,-1,0,0}+28 G_{-1,0,-1,0}
   +4 G_{-1,0,1,0}-32 G_{-1,1,0,0}
\nonumber\\&
   + 10 G_{0,-1,-1,0}-8 G_{0,-1,0,1}+14 G_{0,-1,1,0}+14 G_{0,1,-1,0}
   - 8 G_{0,1,0,1}+6 G_{0,1,1,0}-32 G_{1,-1,0,0}
\nonumber\\&
   + 28 G_{1,0,-1,0}+12 G_{1,0,1,0}-64 G_{1,1,0,0}-11 G_{0,-1} \zeta_{2}
   - 10 G_{-1} \zeta_{3}-22 G_{1} \zeta_{3}+16 G_{-1,0,0} \ln{2}
\nonumber\\&
-8 G_{0,-1,0} \ln{2}-8 G_{0,1,0} \ln{2}
+16 G_{1,0,0} \ln{2}+4 G_{0} \zeta_{2} \ln{2}+8 \zeta_{3} \ln{2})
   -90 (21+8 y+66 y^2-8 y^3
\nonumber\\&   
   +45 y^4) G_{0} \zeta_{3}
   -360 (1+y^2) (11+23 y^2) G_{-1,0} \zeta_{2}
   +180 (1-4 y+4 y^2+4 y^3+3 y^4) G_{0,0} \zeta_{2}
\nonumber\\&
   +180 (1+y^2) (9+41 y^2) G_{0,1} \zeta_{2}
   -360 (1+y^2) (5+9 y^2) G_{1,0} \zeta_{2}
   +360 y^2 (1+y^2) (-3 G_{0,0,0,0}
\nonumber\\&   
   -8 G_{0,0,0,1}+8 G_{0,0,0} \ln{2})
   -360 (9+8 y-8 y^2-8 y^3-17 y^4) G_{-1,0,0,0}
   +360 (9+4 y+40 y^2
\nonumber\\&   
   -4 y^3+31 y^4) G_{0,0,-1,0}
   +360 (5+4 y+18 y^2-4 y^3+13 y^4) G_{0,0,1,0}
\nonumber\\&
   -360 (1+y^2) (5+37 y^2) G_{0,1,0,0}
   -360 (1+y^2) (15+13 y^2) G_{1,0,0,0}
\Big]
\\
d_{-2} &=
-\frac{2}{3}
-\frac{2(1+y^2)}{3 (1-y^2)}G_{0}
\\
d_{-1} &=
-\frac{34}{9} 
+\frac{16}{3} G_{1}
   +\frac{2(1+12 y+25 y^2)}{9 (1-y^2)} G_{0}
   -\frac{4(1+y^2)}{3 (1-y^2)} (4 G_{-1,0}-G_{0,0}-4 G_{0,1}+2 G_{1,0}-4 \zeta_{2})
\\
d_0 &=
-\frac{284}{27}
+\frac{10}{3} \ln{2} 
+\frac{272}{9} G_{1}
-\frac{128}{3} G_{1,1}
+\frac{2}{27 (1-y^2)}
\Big[
   2 (55+129 y+259 y^2) G_{0}
\nonumber\\&
   +3 (43-93 y-173 y^2) \zeta_{2}
   +45 (1+y^2) G_{0} \ln{2}
   +6 (49+39 y+49 y^2) G_{-1,0}
\nonumber\\&   
   -24 (1+12 y+25 y^2) G_{0,1}
   +6 (53+18 y-43 y^2) G_{1,0}
   -18 (1+y^2) (28 G_{-1,-1,0}-4 G_{-1,0,0}
\nonumber\\&   
   -32 G_{-1,0,1}
   +12 G_{-1,1,0}-20 G_{0,-1,0}+2 G_{0,0,0}+8 G_{0,0,1}-22 G_{0,1,0}
   +32 G_{0,1,1}+12 G_{1,-1,0}
\nonumber\\&   
    +2 G_{1,0,0}-16 G_{1,0,1}-30 G_{-1} \zeta_{2}
    +2 G_{0} \zeta_{2}-10 G_{1} \zeta_{2}-13 \zeta_{3})
\Big]
+\frac{2(10-37 y-23 y^2+14 y^3)}{9 (1-y) (1-y^2)}G_{0,0}\,.
\eal
Similarly to $\bI_{1}$, in \eqn{eq:I20} we have expanded a factor 
of $1/C(\ep)^2$ coming from the definition of the renormalized coupling, 
thereby canceling terms of $\gamma_E$ and $\ln(4\pi)$. Switching to 
the standard \MSbar{} convention would imply
\beq
b_0^{\MSbar} = b_0 + \zeta_{2} b_{-2}\,,
\qquad
c_0^{\MSbar} = c_0 + \zeta_{2} c_{-2}
\qquad\mbox{and}\qquad
d_0^{\MSbar} = d_0 + \zeta_{2} d_{-2}\,,
\eeq
with the rest of the expansion coefficients unchanged.

Let us make two comments about the $\bI_2$ operator. First, $\bI_2$ 
corresponds to the sum of all integrated counterterms which involve 
the tree-level factorized matrix element $\SME{Q\bar{Q}}{(0)}{}$, 
except for the renormalization term of the one-loop soft current
which, as noted above, we keep explicit. Second, although the Laurent 
expansion of $\bI_2$ starts at $1/\ep^2$, individual contributions 
to this operator involve qubic poles. These poles come from on the 
one hand double real configurations where the two gluons are collinear 
and both soft. Similarly, the double poles of the real-virtual 
contribution are proportional to the tree level three-parton 
matrix element that develops an extra pole upon integration over 
the soft region of phase space. Given that the double virtual 
matrix element is free of triple poles, these must cancel upon 
combining all integrated subtraction terms.

The above formulae, together with the subtraction terms given in 
the previous sections, complete the full set of ingredients of our 
subtraction scheme. The implementation of the entire procedure 
in a numeric code is then straightforward.

\section{Example: Higgs boson decay to massive bottom quarks}
\label{sec:Hbb}

As stated in the Introduction, the construction given above 
can be applied to compute fully differential NNLO QCD corrections 
to any process with a colourless initial state decaying into a massive 
quark-antiquark pair at leading order. As an illustrative example, 
in this section we report on such a computation for a Standard Model 
Higgs boson decaying into a pair of massive bottom quarks.

We recall that the necessary two-loop currents have been computed in 
\refr{Bernreuther:2004ih,Bernreuther:2004th,Bernreuther:2005gw,Bernreuther:2005rw} 
and \refr{Ablinger:2017hst}. We have verified the correctness of our 
implementations of these formulae by checking the exact agreement among 
them, after accounting for the different conventions. The three parton 
one-loop and four parton tree level matrix elements were obtained with 
a straightforward, direct Feynman-diagram calculation and cross-checked 
with {\tt GoSam} \cite{Cullen:2011ac,Cullen:2014yla}.

To validate our construction, we start by examining the total decay 
width, corresponding to $J=1$, evaluated in the on-shell renormalization 
scheme for the heavy quark,
\beq
\tgam{}_{b\bb} = \tgam{LO}_{b\bb}
    \left[1 
    + \frac{\as}{\pi} \gam^{(1)}_{b\bb}
    + \left(\frac{\as}{\pi}\right)^2 \gam^{(2)}_{b\bb}
    + \Oa{3}
    \right]\,.
\eeq
The NLO correction given by $\gam^{(1)}_{b\bb}$ has been known 
analytically for a long time \cite{Braaten:1980yq,Drees:1990dq}, 
while $\gam^{(2)}_{b\bb}$ has been computed as a series expansion 
in $(m_b^2/m_H^2)$ up to the fourth power \cite{Harlander:1997xa}. 
It has also been calculated exactly for physical values of the 
bottom quark and Higgs boson masses recently \cite{Bernreuther:2018ynm,
Behring:2019oci} and we find perfect agreement with these results. 

\begin{figure}
    \centering
    \includegraphics[scale=0.8]{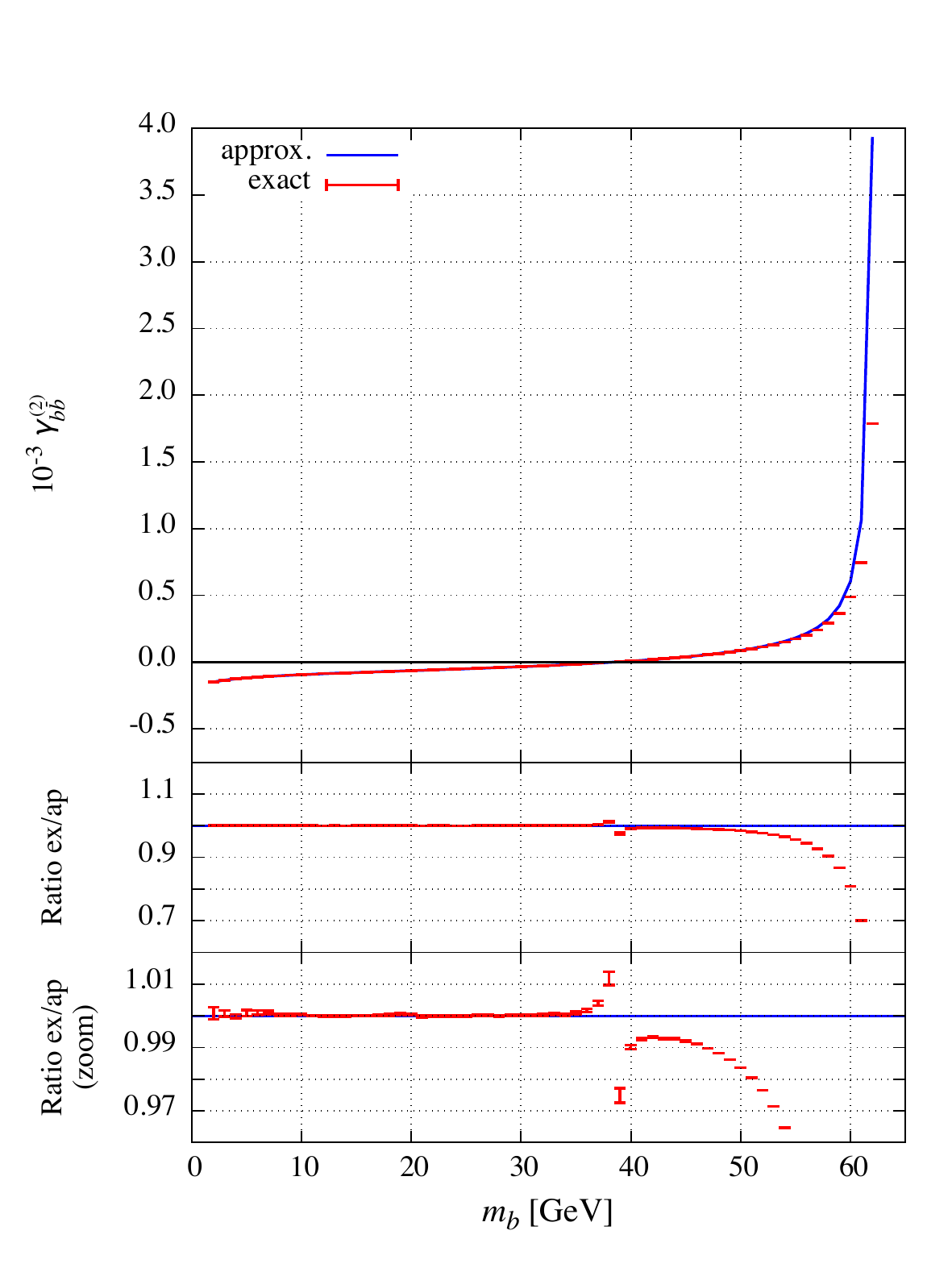}
    \caption{The exact (red) and approximate (blue) NNLO 
    correction $\gamma_{b\bb}^{(2)}$ to the total decay rate 
    of a Standard Model Higgs boson into a heavy quark-antiquark pair 
    as a function of the heavy quark mass. 
    The Higgs boson mass is fixed to its physical value of 
    $m_H = 125.09\,\mbox{GeV}$. The bottom panels show the ratio 
    of the calculations in two different magnification scales.}
    \label{fig:gamma2}
\end{figure}

In order to investigate the validity of the approximate formula for 
the NNLO correction to the total decay width for values of heavy quark masses 
approaching the kinematic threshold, in \fig{fig:gamma2} we compare it to 
the exact computation.
The upper panel shows the value of $\gam^{(2)}_{b\bb}$ as a function 
of the heavy quark mass $m_b$, with the Higgs boson mass fixed to its 
physical value of $m_H=125.09\,\mbox{GeV}$. In order to better appreciate 
the level of agreement, in the lower panels we present the ratio of the 
exact result to the approximate one. We observe that up to around 
$38\,\mbox{GeV}$ (near the threshold for the production of four heavy 
quarks), the agreement is well within 1\%. The reason for the discontinuity 
observed in the ratio for $38\,\mbox{GeV} < m_b < 39\,\mbox{GeV}$ is 
simply due to the fact that the exact and approximate results vanish for 
slightly different values of the heavy quark mass. Between $40\,\mbox{GeV}$ 
and $46\,\mbox{GeV}$, the difference between the two results is still 
below 1\%. For larger values of the heavy quark mass approaching the 
threshold, an all-order expansion in $(m_b^2/m_H^2)$ would be needed, 
that is indeed provided by the exact result.

\begin{figure}
    \centering
    \includegraphics[scale=0.8]{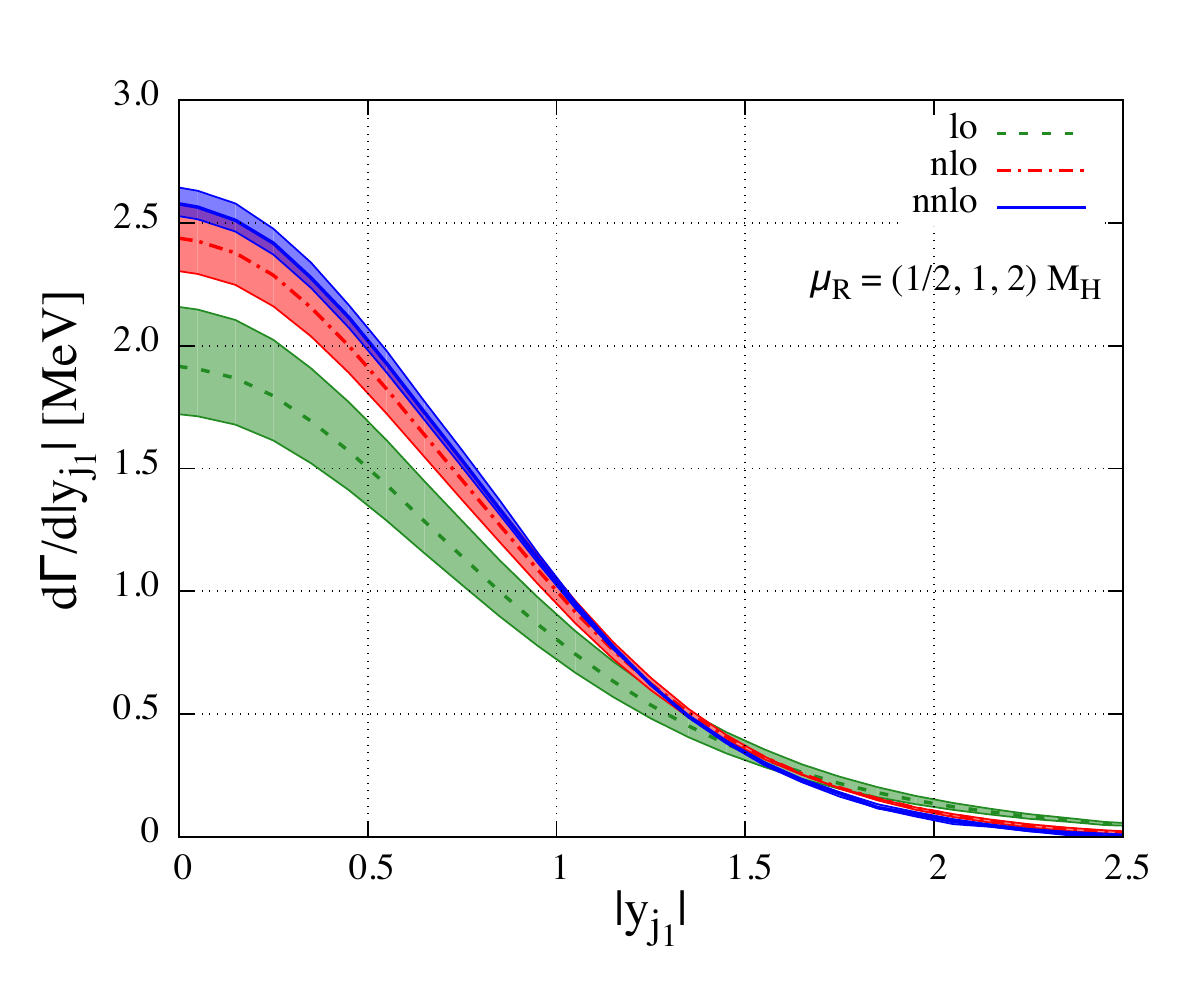}
    \caption{The distribution of the absolute value of the 
    rapidity $|y_{j_1}|$ of the most energetic jet at LO (green),
    NLO (red) and NNLO (blue) accuracy. The bands correspond 
    to the variation of the 
    renormalization scale in the range $\mu_R \in [m_H/2,2m_H]$. 
    Jets are clustered with the Durham algorithm and the
    resolution parameter is fixed at $y_{\mathrm{cut}} = 0.1$.}
    \label{fig:y3b}
\end{figure}

We illustrate the computation of a differential quantity by 
clustering the partons in the final state into jets with the 
Durham algorithm \cite{Catani:1992ua} with the resolution variable 
set to $y_{\mathrm{cut}}=0.1$. In \fig{fig:y3b}, we present the 
differential decay rate in the \MSbar{} scheme with respect to 
the rapidity of the most energetic jet. As can be seen on the 
figure, this variable has a non-singular distribution already 
at LO and so genuine NNLO corrections contribute bin by bin. The 
results presented in \fig{fig:y3b} were obtained with the following 
setup. The Higgs boson mass was set to $m_H = 125.09\,\mbox{GeV}$, 
while the on-shell bottom quark mass was $m_b=4.78\,\mbox{GeV}$, 
which corresponds to $\overline{m}_b(m_H) \simeq 2.79\,\mbox{GeV}$ 
in the \MSbar{} scheme using two-loop running. The strong coupling 
at the relevant renormalization scale was evolved using three-loop 
running starting form $\as(M_Z) = 0.118$.

We recall that the relation between results computed in the on-shell 
and \MSbar{} schemes (denoted with a bar) is given as follows,
\beq
\btgam{}_{b\bb}[J] = \btgam{LO}_{b\bb}[J]
    + \btgam{NLO}_{b\bb}[J]
    + \btgam{NNLO}_{b\bb}[J]
    + \Oa{3}\,,
\eeq
where
\bal
\btgam{LO}_{b\bb}[J] &= \frac{\overline{y}_b^2(\mu_R)}{y_b^2}
    \tgam{LO}_{b\bb}[J]\,,
\\
\btgam{NLO}_{b\bb}[J] &= \frac{\overline{y}_b^2(\mu_R)}{y_b^2}
    \bigg\{\tgam{NLO}_{b\bb}[J] 
    + r_1 \frac{\as(\mu_R)}{\pi} \tgam{LO}_{b\bb}[J]\bigg\}\,,
\\
\btgam{NNLO}_{b\bb}[J] &= \frac{\overline{y}_b^2(\mu_R)}{y_b^2}
    \bigg\{\tgam{NNLO}_{b\bb}[J]
     + r_1 \frac{\as(\mu_R)}{\pi} \tgam{NLO}_{b\bb}[J]
     + r_2 \left(\frac{\as(\mu_R)}{\pi}\right)^2 \tgam{LO}_{b\bb}[J] 
     \bigg\}\,.
\eal
Furthermore \cite{Gray:1990yh,Broadhurst:1991fy,Fleischer:1998dw,Melnikov:2000qh}
\beq
r_1 = -2d_1
\qquad\mbox{and}\qquad
r_2 = 3d_1^2 - 2d_2
\eeq
with
\bal
d_1 &= -\CF \left(1 + \frac{3}{4} L\right)\,,
\\
d_2 &= \CF^2 \left(\frac{7}{128} - \frac{3}{4}\zeta_{3} 
    + 3 \ln{2} \zeta_{2} - \frac{15}{8}\zeta_{2} 
    + \frac{21}{32} L + \frac{9}{32} L^2\right) 
\nonumber\\&    
    + \CA \CF \left(-\frac{1111}{384} + \frac{3}{8} \zeta_{3} 
    + \frac{1}{2} \zeta_{2} - \frac{3}{2} \ln{2} \zeta_{2} 
    - \frac{185}{96} L - \frac{11}{32} L^2\right) 
\nonumber\\&    
    + \CF \TR n_l \left(\frac{71}{96} + \frac{1}{2} \zeta_{2} 
    + \frac{13}{24} L + \frac{1}{8} L^2\right) 
    + \CF \TR \left(\frac{143}{96} - \zeta_{2} + \frac{13}{24} L 
    + \frac{1}{8} L^2\right)\,,
\eal
where $L = \ln(\mu_R^2/m_b^2)$. The relation between the Yukawa 
couplings in the two schemes is given by
\beq
y_b^2 = \overline{y}_b^2
    \bigg[1 
     + r_1 \frac{\as(\mu_R)}{\pi} 
     + r_2 \left(\frac{\as(\mu_R)}{\pi}\right)^2 
     + \Oa{3}\bigg]\,.
\eeq
Although in the \MSbar{} scheme the running mass at the scale around 
the Higgs boson mass is significantly reduced with respect to the on-shell 
value, as is customarily done, we prefer to keep the on-shell mass 
in the definition of the kinematics for the outgoing heavy quark momenta 
in order to mimic the effects related to hadronization that will 
produce mesons with masses close to that value.

To exhibit the reduced theoretical uncertainty due to the higher order 
contributions, we vary the renormalization scale around $m_H$ by a 
factor of two in both directions. With the inclusion of the NNLO 
corrections, we observe a nice convergence of the perturbative expansion 
and the corresponding reduction in the leftover theoretical uncertainty 
parametrized by scale variation.

\section{Conclusions}
\label{sec:concl}

In this paper, we have presented a completely local subtraction scheme 
for computing fully differential NNLO corrections to the production 
of a heavy quark-antiquark pair from a colourless initial state. 
Following the CoLoRFulNNLO method, the construction of our subtraction 
terms starts from the known singular limits of tree-level and one-loop 
massive matrix elements supplemented by momentum mappings that enforce 
exact phase space factorization.
However, we furthermore employ a global strategy simplifying 
simultaneously the computation of the integrated counterterms for 
single and iterated single unresolved emission. This strategy 
is quite general and can be applied in principle also for more generic 
processes. We have implemented further simplifications for the 
specific case of heavy quark-antiquark pair production by including certain 
subleading contributions to the general formulae in the double soft 
limit and the single soft limit for the one-loop heavy quark current. 
As a result, we were able to obtain a very compact analytic result 
for the sum of all integrated subtraction terms with a number of terms 
comparable to that of the two-loop virtual amplitude.

Finally, we have shown the application of our method for the case of 
a Standard Model Higgs boson decaying to a heavy quark-antiquark pair. First, 
we have compared our results for the NNLO correction to the inclusive 
decay rate to the approximate formula of \cite{Harlander:1997xa}, based on a series 
expansion in $(m_b^2/m_H^2)$. Varying the heavy quark mass, we find 
excellent agreement up to values of the heavy quark mass where 
higher order effects in the mass expansion can no longer be neglected. 
Furthermore, as an illustrative example of a differential calculation, 
we have presented the leading jet rapidity distribution at NNLO accuracy.

We conclude by remarking that the present paper contains in full detail 
all formulae that are needed to reproduce the results discussed above, 
and to extend the computation to other heavy quark-antiquark pair production 
processes from colourless initial states.

\section*{Acknowledgments}
We thank Roberto Bonciani, Luca Buonocore, Vittorio Del Duca 
and Pierpaolo Mastrolia for helpful discussions. 
Also we thank Werner Bernreuther, Christian Bogner and 
Oliver Dekkers for kindly providing us the expressions for 
the master integrals computed in \cite{Bernreuther:2013uma}.
This work was supported by grant K 125105 of the National 
Research, Development and Innovation Fund in Hungary.
The research of FT is supported by INFN.

\bibliographystyle{JHEP}
\bibliography{biblio}

\end{document}